\documentclass[apj]{emulateapj}
\usepackage{graphicx,color}
\usepackage[usenames,dvipsnames]{xcolor} % colour
\usepackage{amssymb,amsmath,mathtools,mathrsfs,bm} % maths
\usepackage{epsfig,subfigure,placeins,float} % plots
\usepackage{booktabs,longtable,ctable,multirow} % tables
\usepackage{exscale,relsize} % scale 
\usepackage[normalem]{ulem} % editing
\usepackage{enumerate}
\usepackage[stable]{footmisc}

\usepackage{hyperref}
\hypersetup{
    colorlinks = true,
    citecolor = {MidnightBlue},
    linkcolor = {BrickRed},
    urlcolor = {BrickRed}
}

\newcommand{\corr}{}

\renewcommand{\d}[0]{\mathbf{d}}
\newcommand{\n}[0]{\mathbf{n}}
\newcommand{\s}[0]{\mathbf{s}}
\renewcommand{\a}[0]{\mathbf{a}}

\newcommand{\N}[0]{\mathbf{N}}

\newcommand{\G}[0]{\mathbf{G}}
\newcommand{\B}[0]{\mathbf{B}}
\newcommand{\T}[0]{\mathbf{T}}
\newcommand{\U}[0]{\mathbf{U}}
\renewcommand{\S}[0]{\mathbf{S}}

\newcommand{\half}{\frac{1}{2}}

\newcommand{\be}{\begin{equation}}
\newcommand{\ee}{\end{equation}}
\newcommand{\bea}{\begin{eqnarray}}
\newcommand{\eea}{\end{eqnarray}}

\begin{document}

\title{A CMB Gibbs sampler for localized secondary anisotropies}

\author{
Philip Bull\altaffilmark{1}$^{,}$\altaffilmark{2},
Ingunn K. Wehus\altaffilmark{3}$^{,}$\altaffilmark{2},
Hans Kristian Eriksen\altaffilmark{1},
Pedro G. Ferreira\altaffilmark{2},
Unni Fuskeland\altaffilmark{1},
Krzysztof M. G\'{o}rski\altaffilmark{3}$^{,}$\altaffilmark{4},
Jeffrey B. Jewell\altaffilmark{3},
}

\email{p.j.bull@astro.uio.no}

\altaffiltext{1}{Institute of Theoretical Astrophysics, University of Oslo, P.O. Box 1029 Blindern, N-0315 Oslo, Norway}

\altaffiltext{2}{Astrophysics, University of Oxford, DWB, Keble Road,
  Oxford OX1 3RH, UK}

\altaffiltext{3}{Jet Propulsion Laboratory, California Institute of
  Technology, Pasadena, CA 91109, USA}

\altaffiltext{4}{Warsaw University Observatory, Aleje Ujazdowskie 4, 00-478 Warszawa, Poland}

%\date{Received - / Accepted -}

\begin{abstract}
As well as primary fluctuations, CMB temperature maps contain a wealth of additional information in the form of secondary anisotropies. Secondary effects that can be identified with individual objects, such as the thermal and kinetic Sunyaev-Zel'dovich (SZ) effects due to galaxy clusters, are difficult to unambiguously disentangle from foreground contamination and the primary CMB {\corr however}. We develop a Bayesian formalism for rigorously characterising anisotropies that are localised on the sky, taking the TSZ and KSZ effects as an example. Using a Gibbs sampling scheme, we are able to efficiently sample from the joint posterior distribution for a multi-component model of the sky with many thousands of correlated physical parameters. The posterior can then be exactly marginalised to estimate properties of the secondary anisotropies, fully taking into account degeneracies with the other signals in the CMB map. We show that this method is computationally tractable using a simple implementation based on the existing Commander component separation code, and also discuss how other types of secondary anisotropy can be accommodated within our framework.
\end{abstract}


%===============================================================================
\section{Introduction} \label{sec:intro}

Observations of the temperature anisotropies of the Cosmic Microwave Background (CMB) radiation have been instrumental in providing high-precision measurements of important cosmological quantities such as the age, geometry, and energy content of the Universe. But while the task of characterising the primary anisotropies may seem essentially complete -- as of the Planck 2013 data release, measurements of the temperature autospectrum of the CMB are cosmic variance dominated for multipoles $\ell \lesssim 1500$ \citep{2013arXiv1303.5075P} -- a wealth of information remains to be picked out of CMB temperature maps in the form of secondary anisotropies.

Secondary anisotropies are distortions of the primary CMB signal due to inhomogeneities between the surface of last scattering and the observer \citep{2008RPPh...71f6902A}. Gravitational effects such as weak lensing and the integrated Sachs-Wolfe (ISW) effect have been strongly detected \citep{2011PhRvL.107b1301D, 2013arXiv1303.5079P, 2013arXiv1303.5077P, Hernandez-Monteagudo:2013vwa}, as have scattering phenomena such as the thermal Sunyaev-Zel'dovich (TSZ) effect \citep{1984Natur.309...34B, 2011ApJ...736...61S, 2012PhRvD..86l2005W, 2013arXiv1303.5081P, 2013arXiv1303.5089P}. The kinetic Sunyaev-Zel'dovich (KSZ) effect, caused by the Doppler boosting of CMB photons scattered off ionised gas travelling with a bulk peculiar velocity, has also recently been detected \citep{2012PhRvL.109d1101H, 2013ApJ...778...52S}. These effects variously probe the spatial and temporal variations of the gravitational potential, the density field, and the peculiar velocity field on large scales, and can therefore furnish tests of dark energy, modified gravity, and even the inflationary epoch by constraining the geometry, expansion and growth history of the Universe.

The secondary anisotropies are generally dominated by foregrounds and the primary CMB, so detecting and characterising them is a delicate process, even with modern high-resolution data. Detections of secondaries can be roughly divided into two categories: {\corr {\it localised} on the sky (compact)}, where anisotropies in a given direction can be identified as being caused by a particular astrophysical object; and {\corr {\it non-localised} (diffuse)}, where a secondary signal due to the combination of many objects is detected statistically across a larger region of the sky.

A number of different methods have been used to measure non-local signals: fitting models to the angular power spectrum \citep{2010ApJ...722.1148F, 2011ApJ...736...61S}; using some other statistical property of the CMB map, such as higher-order (non-Gaussian) moments, to separate off a given signal \citep{2005MNRAS.359..261P, 2012PhRvD..86l2005W, 2013MNRAS.429.1564M}; stacking the signal from many directions to average down all but the target signal \citep{2008ApJ...683L..99G, 2009arXiv0907.0233D, Komatsu:2010fb}; and cross-correlating the CMB map with tracers (e.g. galaxies) that are uncorrelated with all but the target signal \citep{2003ApJ...597L..89F, 2004PhRvD..70h3536A, 2006PhRvD..74f3520G, 2008PhRvD..78d3519H, 2012PhRvL.109d1101H, 2012PhRvD..86h3006S, 2014MNRAS.438.1724H}. These all allow small secondary anisotropy signals to be picked out by essentially combining the signal from all available pixels into a single statistical quantity.

Detecting localised signals is often a more difficult prospect. Because astrophysical objects typically subtend small angles on the sky, only a limited number of pixels are available to provide information about a given object, making it harder to attain a high enough signal-to-noise ratio to get a definitive detection. The limited amount of information also makes it harder to disentangle other signals from the secondary anisotropy, especially if they have similar frequency spectra or shapes/angular sizes, and there are fewer options for averaging-down contaminating signals.

Previous approaches to this problem have tended to rely on a combination of frequency information and angular filters matched to the size/shape of the secondary anisotropy to try and pick-out its signal, while rejecting or at least averaging down other signals as much as possible {\corr \citep{2001A&A...374....1A, 2002MNRAS.336.1057H, 2005MNRAS.356..944H, forni2005adapted, 2006A&A...459..341M, 2007MNRAS.377..253M, 2009MNRAS.398.2049F, 2011ApJ...736..116M, 2012arXiv1211.4345A, 2012MNRAS.427.1384C, Melin:2012iz}}. Such methods may either be blind, applying filters of a range of sizes over the entire map, or can apply parameter estimation techniques to fit parametrised models to the objects, given prior information on their positions and/or sizes. These methods work well if the secondary has a distinctive spectrum and multi-frequency data are available, but this is not always the case -- the KSZ effect has the same flat spectrum as the primary CMB, for example, and only a few CMB experiments have more than one or two frequency bands.

Angular information is also valuable, but most filtering and model fitting techniques are unable to blindly distinguish signals with similar angular structures. As such, the estimated signal for an individual object will retain some level of residual contamination from other fluctuating components. One example of this is the contamination of the KSZ signal by the primary CMB -- primary anisotropies on the arcminute scales characteristic of galaxy clusters cannot be fully removed by a filter, and so will either bias the estimated signal, or must be treated as an effective source of noise, significantly increasing the statistical errors \citep{2001A&A...374....1A, 2005MNRAS.356..944H, 2009MNRAS.398.2049F, 2012MNRAS.427.1384C}.

In this paper we describe a novel method for characterising localised secondary anisotropies, based on applying Bayesian inference to a physical parametric model of all relevant signals on the sky. We use the Gibbs sampling technique to efficiently reconstruct the joint posterior distribution of the full sky model, which typically involves many hundreds of thousands of parameters for realistic datasets. With the posterior in hand, we can then marginalise over all other parameters to produce statistically-robust, unbiased estimates of the properties of the secondary anisotropies.

Importantly, because all of the signals that contribute to the CMB map are explicitly modelled, degeneracies with local fluctuations in other signals can be fully taken into account. Instead of being treated as random noise, the fluctuations are reconstructed from the data, allowing them to be cleanly separated from the secondary signal in a statistical manner. For the primary CMB, this is equivalent to performing a constrained Gaussian realisation of the anisotropies behind the cluster. The uncertainties associated with this procedure are automatically propagated in full by the Gibbs sampling scheme.

The paper is organised as follows. In Section \ref{sec:gibbs} we outline a general Gibbs sampling scheme for estimating localised signals in the presence of primary CMB anisotropies, various types of foreground emission, and noise. We then specialise to a couple of example secondary anisotropies: the TSZ effect (Section \ref{sec:tsz}), and the KSZ effect (Section \ref{sec:ksz}), and demonstrate a simple proof-of-concept implementation of the SZ Gibbs scheme in Section \ref{sec:sampler}. Examples of other localised signals that can be accommodated by our framework are discussed in Section \ref{sec:other}, and we conclude in Section \ref{sec:discussion}.
\vfill

%===============================================================================
\section{Gibbs sampling of localised signals} \label{sec:gibbs}

Gibbs sampling is a popular Monte Carlo technique for performing Bayesian inference on complex parametric models. In this section, we outline a Gibbs sampling scheme for the joint estimation of CMB anisotropies, galactic foregrounds, and spatially-localised signals from multi-frequency full-sky data. This is based on the CMB analysis framework previously described by \cite{Jewell:2002dz, Wandelt:2003uk, 2004ApJS..155..227E, 2008ApJ...676...10E}, which allows for straightforward marginalisation over both CMB and foreground signals by sampling from the joint posterior distribution of all components.

\subsection{Data model} \label{sec:data-model}

We begin by defining a data model for an observation of the sky at a given frequency,
\begin{equation}
%d_{p}(\nu) = B_{p}^{q}(\nu) G^{r}_{q}(\theta;\nu)
%T^{i}_{r} a_{i} + n_p(\nu),
\mathbf{d}(\nu) = \mathbf{B}(\nu) \sum_{i=1}^{N_\textrm{comp}}
\mathbf{G}_i(\nu) \mathbf{T}_{i} \mathbf{a}_{i} +
\mathbf{n}(\nu). \label{eq:datamodel}
\end{equation}
In this expression, $\mathbf{d}$ is a vector of observed values, $d_{p}(\nu)$, 
for each pixel $p$ and
frequency $\nu$, and $\mathbf{n}$ denotes instrumental noise. The
signal components are broken down into a set of unknown stochastic
amplitudes ($a_i$), an amplitude-to-sky projection operator
($T_i$), a frequency-dependent mixing operator
($G_i$), and an instrumental beam convolution operator
($B$). Note that the index $i$ runs over both signal types
(CMB, foregrounds, SZ amplitudes etc.) and individual components
within each signal type (different SZ clusters, CMB harmonics
etc.).

Next, we specify the statistical properties of $a_i$ and $n_p(\nu)$. 
In this paper we assume the signal amplitudes and noise to be
Gaussian, having covariance matrices $\mathbf{S}$ and $\mathbf{N}$ respectively. 
One is often interested in estimating the signal covariance matrix from the 
data, so in general $\mathbf{S}$ is unknown and must be jointly estimated with 
the rest of the model parameters. The basic structure of $\mathbf{S}$ 
can often be specified {\it a priori}, and parametrised in terms of a 
relatively small number of free parameters. For example, an isotropic,
Gaussian CMB component will have a diagonal signal covariance matrix with the 
CMB angular power spectrum coefficients $C_\ell$ along the diagonal. The noise 
covariance matrix will be assumed to be completely known, although in principle 
it could also be specified using some parametric model.

To complete the data model, we must define an inventory of relevant signal 
components and specify the properties of $\mathbf{G}$ and $\mathbf{T}$ for 
each. Depending on the component, these can also be modelled parametrically, 
with free parameters to be estimated from the data; for example, the frequency 
mixing matrix of a galactic synchrotron component might take the form of a 
power-law spectrum with an unknown spectral index. We will make no assumptions 
about the statistical distributions followed by these additional parameters for 
the time being; as will soon become apparent, they are not generally Gaussian.

\subsection{Posterior mapping by Gibbs sampling} \label{sec:posterior-mapping}

Once the data model has been defined, the remaining problem is to map out the 
full joint posterior distribution {\corr $P(\a,\S,\G, \T|\d)$ (where it should be noted that we have implicitly conditioned on the beam, $\B$, and noise covariance, $\N$)}. The posterior is unlikely to take the form of a known analytic distribution that can be sampled from directly, suggesting the use of a Markov Chain Monte Carlo (MCMC) method to obtain samples. Owing to the extremely high dimensionality of the problem (e.g. there are $(\ell_\mathrm{max} + 1)^2$ parameters for the CMB component alone), popular MCMC techniques like Metropolis-Hastings and nested sampling are unsuitable, as they tend to scale poorly with dimension \citep{2014MNRAS.437.3918A}. Maximum likelihood techniques are a possible alternative, although these are by nature approximate and thus fail to fully propagate statistical uncertainty.

\begin{figure}[t]
  \includegraphics[width=\columnwidth]{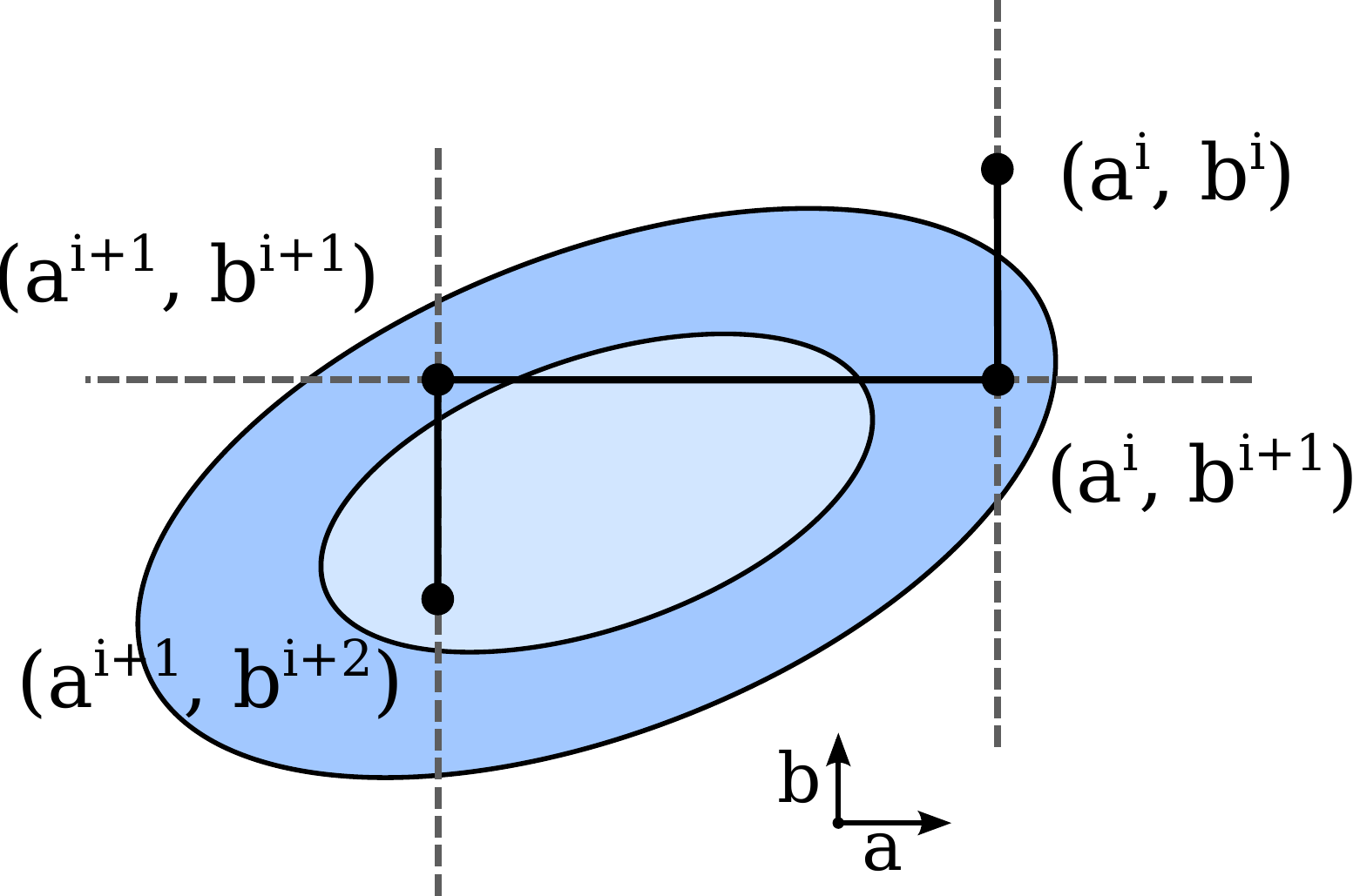}
  \caption{Illustration of the iterative sampling procedure that forms the basis of Gibbs sampling. The algorithm alternately samples from the conditional distributions of the various parameters -- $P(a | b^i)$, then $P(b|a^{i+1})$ and so on. Each sub-step is in an orthogonal direction in parameter space.}
  \label{fig:gibbs}
\end{figure}

Instead, we will make use of the Gibbs sampling algorithm (which is technically a special case of Metropolis-Hastings). While it may not be possible to sample from the joint posterior directly, it is often the case that it can be broken down into a set of conditional distributions that are tractable. One can show that iteratively sampling from the conditionals (Fig. \ref{fig:gibbs}) results in a set of samples that eventually converges to the joint posterior. In other words, by breaking the sampling problem up into a series of comparatively simpler steps, we can reconstruct the full posterior distribution without recourse to approximations or any other ``lossy'' procedures. This holds even for extremely high-dimensional parameter spaces if there are high-dimensional joint conditionals that can be evaluated efficiently (for example, if most parameters can be drawn from a multivariate Gaussian distribution).

As shown in Fig. \ref{fig:gibbs}, Gibbs samplers explore the parameter space using a series of orthogonal sub-steps. This is inefficient for parameters that are strongly correlated, for which the optimum strategy would be to explore along the degeneracy direction. As such, care must be taken to either avoid parametrisations with degenerate parameters, or to write down joint conditionals for strongly correlated parameters that can be evaluated directly. Otherwise, the chain will spend a long time slowly exploring the strongly correlated subspace, and the resulting MCMC chain will have a long correlation length, resulting in fewer independent samples.

For our problem, the Gibbs scheme is
\begin{align}
\label{eq:a_samp}
\a^{i+1} &\leftarrow P(\a|\S^i,\G^i,\T^i,\d) 
\\\label{eq:S_samp}
\S^{i+1} &\leftarrow P(\S|\a^{i+1},\G^i,\T^i,\d) 
\\\label{eq:G_samp}
\G^{i+1} &\leftarrow P(\G|\a^{i+1},\S^{i+1},\T^i,\d) 
\\\label{eq:T_samp}
\T^{i+1} &\leftarrow P(\T|\a^{i+1},\S^{i+1},\G^{i+1},\d).
\end{align}
One can of course further subdivide the conditional sampling steps using Bayes' Theorem and other basic statistical relations if necessary. In general, though, it is more efficient to simultaneously sample as many parameters as possible in each step, in order to reduce the correlation length of the chain.

For the remainder of this section, we will show how each of these Gibbs steps can be sampled {\corr in practice}. For a more detailed discussion of the general properties of Gibbs sampling techniques, see \cite{doi:10.1080/01621459.1990.10476213, AmStat1992}.

\paragraph{Joint amplitude sampling}
Under the assumption that the likelihood is Gaussian, Eq. (\ref{eq:a_samp}) reduces to a single multivariate Gaussian distribution for all $\a$, meaning that the amplitude parameters for all components (potentially hundreds of thousands of them) can be sampled simultaneously. Simplifying the notation for the signal to $\s=\U\cdot\a$, with $\U=\B\G\T$, one can see this by writing
\begin{eqnarray}
P(\a|\d,\S,\G,\T) &\propto& P(\d|\a,\S,\G,\T)P(\a|\S,\G,\T) \nonumber\\
&\propto& e^{-\half (\d-\U\cdot\a)^T \N^{-1}  (\d-\U\cdot\a) }
\cdot e^{-\half \a^T \S^{-1} \a} \nonumber\\
&\propto& e^{ -\half (\a-\hat\d)^T \left(\S^{-1} + \U^T \N^{-1} \U \right)  (\a-\hat\d)}.\label{eq:lin_amp}
\end{eqnarray}
The distribution has covariance $\left(\S^{-1} + \U^T \N^{-1}\U \right)^{-1}$ and (Wiener-filtered) mean
\begin{align}
\hat\d &= \left( \S^{-1} + \U^T \N^{-1} \U \right)^{-1} \U^T \N^{-1} \d.
\end{align}
We have suppressed sums over frequency here; see \cite{2008ApJ...676...10E} for a derivation of the above in the full multi-frequency case.

Sampling from this distribution is conceptually straightforward: one first generates a pair of vectors of $\mathcal{N}(0,1)$ random variables, ($\omega_0, \omega_1$), and then solves the linear system $\mathbf{M} \a = \mathbf{b}$ for $\a$, where
\begin{eqnarray}
\mathbf{M} &=& \S^{-1}+\U^T\N^{-1}\U \label{eqn-linsys-M} \\
\mathbf{b} &=& \U^T\N^{-1}{\d}+\S^{-\half}\omega_0+(\U^T\N^{-1}\U)^{\half}\omega_1. \label{eqn-linsys-b}
\end{eqnarray}
{\corr In practice}, solving this system is a significant computational challenge, owing to its high dimensionality and typically poor conditioning of the matrix operator $\mathbf{M}$. We discuss this more in Section \ref{sec:crsolver}; see also \cite{2008ApJ...676...10E} for a detailed discussion of this problem.

A useful feature of the joint amplitude sampling step is marked by the presence of prior-dependent ($\S$) terms in Eqs. (\ref{eqn-linsys-M}) and (\ref{eqn-linsys-b}). These ensure that the solution is defined even in regions where the data have been masked. Solving the linear system therefore amounts to drawing a constrained realization of the amplitudes that is statistically consistent with the available data and other parameters of the data model. The availability of `uncut' amplitude map samples simplifies subsequent Gibbs steps that may rely on spherical harmonic analysis, such as those involving angular power spectrum estimation.

\paragraph{Signal covariance}
Eq. (\ref{eq:S_samp}) can be written as
\begin{eqnarray}
P(\S|\d,\G,\a) &\propto& P(\d|\S,\G,\a)P(\S|\G,\a) \propto P(\S|\G,\a) \nonumber\\
&=& P(\a|\G,\S)\cdot\frac{P(\S|\G)}{P(\a|\G)} \propto P(\a|\G,\S) \nonumber\\
&=& \frac{e^{-\half \a^T \S^{-1} \a}}{\sqrt{|\S|}}. \label{eqn-S-dist}
\end{eqnarray}
This is an inverse Wishart distribution for $\S$, assuming flat priors for 
$\S$ and $\a$. Prior knowledge of the form of $\S$ can be used to further 
simplify Eq. (\ref{eqn-S-dist}) to one of the special cases of the inverse 
Wishart distribution.

Notice that the dependence on the data, $\d$, has dropped out of Eq. (\ref{eqn-S-dist}). This is because we are now conditioning on $\a$, which contains all of the information necessary to estimate $\S$.

\paragraph{General parameters}

For Eqs. (\ref{eq:G_samp}) and (\ref{eq:T_samp}), a simple application of 
Bayes' Theorem yields
\begin{eqnarray}
P(\G|\d,\a,\S,\T) &\propto& P(\d|\G,\a,\S,\T)P(\G|\a,\S,\T) \nonumber\\
 &\propto& e^{-\half (\d-\U\cdot\a)^T \N^{-1}  (\d-\U\cdot\a) } \cdot P(\G)  \label{sed-pdf}\\
P(\T|\d,\a,\S,\G) &\propto& P(\d|\T,\a,\S,\G)P(\T|\a,\S,\G) \nonumber\\
&\propto& e^{-\half (\d-\U\cdot\a)^T \N^{-1}  (\d-\U\cdot\a) } \cdot P(\T). \label{eqn-shape-pdf}
\end{eqnarray}
In this very general notation, the above equations do not tell us how
to sample from these distributions, and the priors for $\G$ and $\T$ are left
general. This is because the form of these terms depends on the particular 
model chosen for each component. In the following sections we will consider 
specific examples (summarised in Table \ref{sec:crsolver}), for which the above equations simplify significantly.

\begin{table*}
 \centering{
 {\renewcommand{\arraystretch}{1.8} 
 \begin{tabular}{|l|c|c|c|c|}
 \hline
 {\bf Component Type} & {\bf Spectrum} $\mathbf{G}(\nu)$ & {\bf Spatial dependence} $\mathbf{T}$ & {\bf Covariance} $\mathbf{S}$ & {\bf Amplitude} $\mathbf{a}$ \\
 \hline
  CMB anisotropies & $1$ & $Y_{\ell m} (p)$ & $C_\ell \delta_{\ell\ell'}\delta_{mm'}$ & $a_{\ell m}$ \\
  %Spatial template (fixed spectrum) & $f(\nu)$ & Pre-defined template $T(p)$ &  $\infty$ & $A$ \\
  Spatial template (e.g. monopole/dipole) & $\delta_{\nu \nu^\prime}$ & Pre-defined template $T(p)$ & $\infty$ & $A_{\nu^\prime}$ \\
  Pixel-based foreground & $f(\nu; \bm{\theta}_p)$ & $\delta_{p p^\prime}$ & $\infty$ & $A_{p^\prime}$ \\
  Thermal SZ (clusters) & $x \frac{e^x - 1}{e^x + 1} - 4$ & $T_\mathrm{TSZ}(p, \hat{n}_j, \bm{\theta}_j)$ & $\infty$ & $A_j$ \\
  Kinetic SZ (clusters) & $1$ & $T_\mathrm{KSZ}(p, \hat{n}_j, \bm{\theta}_j)$ & See Eq. (\ref{eqn-vpec-correlation}) & $v^{\mathrm{los}}_j$ \\
 \hline
 \end{tabular} } }
 \caption{Signal component types, defined by their spectral and spatial dependence.\label{tbl-components} \label{tbl:components}}
\end{table*}

\subsection{CMB component} \label{sec:cmb}
% CMB component

The CMB signal, ${\bf s}$, is a statistically isotropic field that can be expanded in spherical harmonics, $Y_{\ell m}({\hat n})$ (where ${\hat n}$ is a unit vector direction in the sky), such that the signal in an individual pixel $p$ is
\begin{eqnarray}
{\bf s}_p=\sum_{\ell m} a_{\ell m}Y_{\ell m}({\hat n}_p).
\end{eqnarray}
We identify the coefficients $a_{\ell m}$ and spherical harmonic operator $Y_{\ell m}({\hat n}_p)$ with the amplitudes ($\a$) and projection operator ($\T$) for this component respectively. The signal covariance ($\S$) is given in harmonic space by
\begin{eqnarray}
\langle a_{\ell m}a_{\ell'm'}\rangle\equiv S =C_\ell \delta_{\ell\ell'}\delta_{mm'}, \nonumber
\end{eqnarray}
which reduces Eq. (\ref{eqn-S-dist}) to a set of inverse gamma distributions, independent for each $\ell$. The mixing operator ($\G$) is the identity, since the CMB frequency spectrum is very close to blackbody, and so has a flat spectrum in brightness temperature.

\subsection{Extended foreground components and offset estimation} \label{sec:foregrounds}
% Different types of foreground

Galactic synchrotron, free-free, thermal dust emission, and other extended foregrounds typically have complex spatial structures that do not follow simple statistical distributions like the CMB. As such, it is critical to include in the analysis some frequency channels for which these signals dominate -- say, below 30 GHz for synchrotron/free-free or above 353 GHz for dust. It is then straightforward to reconstruct these components pixel-by-pixel, although a notable exception is spinning dust, which does not dominate at any frequency \citep{2014A&A...565A.103P}, and is consequently subject to considerable degeneracies.

For each component one must write down an explicit parametrisation of $\mathbf{G}(\nu)$, based on some small number of parameters, $\theta_p$, per  pixel. For example, synchrotron is often modelled in terms of a power law in brightness temperature, $f_{\textrm{synch}}(\nu; \beta_{\textrm{s}}) = \nu^{\beta_{\textrm{s}}}$, while thermal dust is well described by a modified blackbody with free emissivity index and temperature. These parameters may then be sampled using Eq. (\ref{sed-pdf}), which reduces to an effective $\chi^2$ mapping of the respective parameters. 

In addition to foreground parameters, there are significant uncertainties in the absolute offset and dipole terms of a given CMB map. These degrees of freedom are easily described in terms of four full-sky templates: one full-sky constant and three orthogonal dipole modes, each with an unconstrained overall linear amplitude. The appropriate sampling algorithm in this case is the usual Gaussian given in Eq. (\ref{eq:lin_amp}), with $\mathbf{G}(\nu,\nu')=\delta_{\nu\nu'}$, $\mathbf{S}^{-1}=0$, and $\mathbf{T}$ listing the four monopole and dipole templates.

\subsection{Spatially-localised components} \label{sec:localised}

As well as separating foregrounds and other effects from the primary CMB, we are also interested in detecting and characterising secondary anisotropies. We will concentrate on the thermal and kinetic Sunyaev-Zel'dovich effects due to galaxy clusters in subsequent sections, but for now the discussion is kept general.

Unlike extended foreground components, which are typically modelled as large coherent structures covering a sizeable portion of the sky, many secondary anisotropies are associated with discrete objects, and are therefore strongly localised. The spatial distribution of the secondary anisotropy is best captured by specifying a collection of spatial templates of limited size, each centred around the location of an individual object. Each localised template will have a separate amplitude associated with it, although the amplitudes may be correlated between objects. The shapes and frequency spectra of the templates may vary from object to object too, so to account for this one can define parametric spatial and spectral profiles with parameters that can be tuned (or sampled) for each object individually.

To define the model for this type of component, one must also specify the number and positions of its constituent objects, and basic information on the template for each object, such as its angular size. This requires a source of prior information, typically in the form of a catalogue of objects. For SZ clusters, for example, one could use a catalogue of `candidate' clusters from a blind SZ detection algorithm (Appendix \ref{app:blind}), or one of a number of X-ray cluster catalogues. Clearly, the specification of the component will only be as complete (and accurate) as the catalogue used, leading to issues with missing or duplicated objects, position errors, and the like. We will return to these problems later.

An object with index $j$, centred on direction $\hat{n}_j$, has the projection operator
\begin{equation}
\T_j = T_i(p, \hat{n}_j, \bm{\theta}^T_j),
\end{equation}
where $T_i$ is a parametric spatial profile shared by all objects of this type of component, and $\bm{\theta}^T_j$ are the profile parameters for the individual object. For localised signals, $T_i$ will typically be zero beyond some given angular distance from $\hat{n}_j$, although this is not compulsory.

Similarly, the frequency mixing operator is given by
\begin{equation}
\G_j = f_i(\nu, \bm{\theta}^G_j),
\end{equation}
where $f_i$ is a shared parametric spectral function and $\bm{\theta}^G_j$ are the spectral parameters for an individual object. Each object has a single overall amplitude, $a_j$. Correlations between amplitudes are specified by a single signal covariance matrix for {\it all} objects, i.e.
\begin{equation}
\S_i = \langle a_j a_k \rangle.
\end{equation}

{\corr In the subsequent sections}, we will consider the TSZ and KSZ effects for galaxy clusters as two specific examples of localised components.

{\corr
\subsection{Instrumental properties} \label{sec:beamnoise}

Finally, we note the ability of our framework to account for uncertainty in the characterisation of the instrument. Eq.~(\ref{eq:datamodel}), and many of the expressions that follow, have been derived under the assumptions of Gaussian instrumental noise, linear beam convolution, and correct gain calibration. These assumptions are reasonable for real instruments such as Planck, although subject to some complications. The gain calibration is typically uncertain, but can be accounted for by multiplying (\ref{eq:datamodel}) by an additional (constant) parameter per frequency channel and marginalising over it.

The noise is typically well-approximated as Gaussian, but can be correlated between neighbouring pixels (e.g. for Planck). The generalised framework presented above already accounts for correlated Gaussian noise, but existing implementations of the joint amplitude sampling step tend to assume uncorrelated noise; the presence of off-diagonal components of the pixel-space noise covariance matrix can significantly increase the computational complexity of solving the linear system, making it difficult to solve except for with low-resolution pixelisations. There is no reason why specialised linear solvers could not be employed to make this more efficient for finer pixelisations, however.

Similarly, one typically assumes symmetric beams that are independent of frequency \citep{2004ApJS..155..227E}, and constant with respect to position on the sky. Again, this is not a necessary condition of our general framework, but is used to increase the efficiency of the amplitude sampling step, which relies on many evaluations of the beam-convolved signal model -- factoring the beam out as a constant can considerably reduce the computational complexity of solving the linear system. The need for frequency-independent (matched) beams has recently been relaxed in a computationally-efficient linear solver \citep{2014ApJS..210...24S}, which could in principle handle asymmetric beams too. Position-dependent beams add an extra layer of complexity, however. Marginalisation over uncertainties in symmetric beam profiles has been demonstrated within a Gibbs sampling framework by sampling the coefficients of an eigenmode expansion of the beam profile \citep[e.g.][]{2013arXiv1303.5075P}.

}

%===============================================================================
\section{Thermal SZ from galaxy clusters} \label{sec:tsz}

The thermal SZ effect is caused by the Compton scattering of CMB photons by hot gas in the intergalactic medium \citep{1972CoASP...4..173S}. The CMB gains energy from the gas, effectively leading to a shift in its spectrum along the affected line of sight. This is manifested as an apparent decrement in the CMB temperature at low frequencies ($\nu \lesssim 217$ GHz), and an increment at higher frequencies, which distinguishes TSZ from the flat-spectrum primary CMB signal. As free electrons dominate the scattering, the magnitude of the shift depends primarily on the integrated electron pressure along the line of sight. In galaxy clusters, the thermal pressure can be related to the cluster size and mass, and so the TSZ effect can be used as a way of probing a cluster's physical properties. This is useful for understanding how structure forms, as well as providing constraints on cosmological parameters such as the normalisation of the matter power spectrum {\corr \citep{Battye:2003bm, Allen:2011zs}}. Another useful property of the TSZ effect is that the surface brightness is constant as a function of redshift \citep{1972CoASP...4..173S, 1995ARA&A..33..541R}, making it possible to detect clusters out to high redshift ($z \gtrsim 1$).

We will assume here that a catalogue of positions, angular sizes, and redshifts of clusters has already been compiled from a previous blind survey, so that our task is to accurately characterise the clusters' properties. This is particularly critical for the most massive clusters, as some cosmological tests are extremely sensitive to the location of the high-mass cut-off of the cluster mass function \citep{2000ApJ...541...10M}. Note that while blind detection algorithms are capable of providing some information on cluster properties, they often rely on simplified or approximate treatments of issues such as residual foreground contamination, statistical error propagation, overlap between clusters, and so on, so it is important to perform a more specialised characterisation post-detection.

\subsection{Model definition} \label{sec:tsz-model}

Following the discussion in Section \ref{sec:localised}, we begin by defining a parametric spatial template for the cluster TSZ signal. The fractional temperature change due to the thermal SZ effect along a line of sight is given by \citep{1972CoASP...4..173S}
\bea \label{eqn-tsz-dt}
\frac{\Delta T}{T} &=& f(\nu) y(\hat{n}) \\
y(\hat{n}) &=& \frac{\sigma_T}{m_e c^2} \int P_e(\hat{n}, l) dl,
\eea
where $P_e$ is the electron pressure, $l$ is a distance along the line of sight, and $f(\nu)$ is the frequency spectrum,
\be
f(\nu) = x \frac{e^x - 1}{e^x + 1} - 4; \;\;\; x = h\nu / k_B T_\mathrm{CMB}, \label{eqn-tsz-spectrum}
\ee
which can immediately be identified with the mixing operator, $\G$. In our model, we adopt the `universal' pressure profile of \cite{2010A&A...517A..92A},
\be
P_e(x) = \bar{P}_{500}(z) \frac{P_0 (M_{500} / M_\star)^{\alpha_p(x)}}{(c_{500}
  x)^\gamma [1 + ({c_{500} x)}^\alpha ]^{\frac{\beta - \gamma}{\alpha}}}, \label{eqn-pe}
\ee
where $x = r/R_{500}$ and $M_\star = 3\times10^{14} h^{-1} M_\odot$. The pressure at radius $R_{500}$ in a gravity-only self-similar model is \citep{2007ApJ...668....1N}
\be
\bar{P}_{500}(z) = 1.65\times10^{-3}h(z)^\frac{8}{3} (M_{500} / M_\star)^\frac{2}{3} h^2 ~\mathrm{keV cm}^{-3},\nonumber
\ee
and the running of the mass scaling with radius is well-fit by $\alpha_p(x) = 0.22\; ( 1 - 8x^2/(1 + 8x^3))$. The normalisation and shape parameters of the universal profile have best-fit values of $[P_0,c_{500}, \gamma, \alpha, \beta] = [8.403 h^{-\frac{3}{2}}, 1.177, 0.3081, 1.0510, 5.4905]$, calibrated from the REXCESS sample of 33 local X-ray clusters at small radii \citep{2007A&A...469..363B} and hydrodynamic simulations at large radii.

\begin{figure}[t]
  \hspace{-1em}\includegraphics[width=1.05\columnwidth]{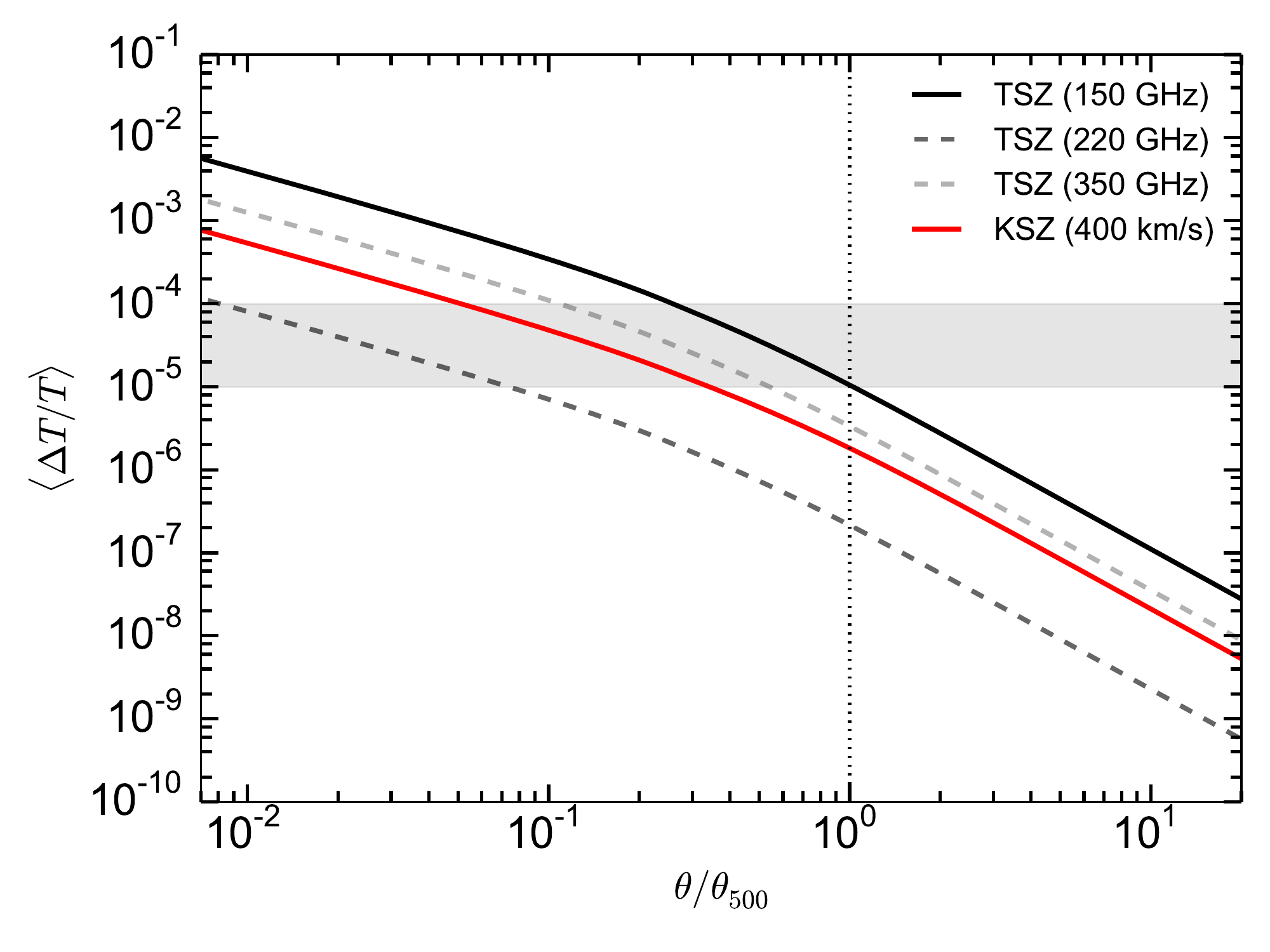}
  \caption{Temperature fluctuation due to thermal SZ (black/grey) and kinetic SZ (red) at 150, 220, and 350 GHz, averaged over a circular top-hat aperture of radius $\theta$. The lines shown are for a `typical' SZ cluster at $z=0.1$, with $M_{500} = 10^{14} M_\odot$, $R_{500} = 1$ Mpc, and line of sight velocity $\bm{v} \cdot \hat{n} = 400$ kms$^{-1}$. The grey band illustrates the approximate range of temperature fluctuations for the primary CMB. We have defined $\theta_{500} = R_{500} / D_A(z)$.}\vspace{0.8em}
  \label{fig:tsz-ksz-profile}
\end{figure}

The pressure profile is fully specified once the characteristic mass, radius,
and redshift of the cluster are given. The TSZ projection operator for a cluster $j$, with pressure profile centred about the direction $\hat{n}$, is then
\be
T^\mathrm{TSZ}_j(p) = y(p, \hat{n}_j, \bm{\theta}^T_j), \label{eqn-tsz-template}
\ee
where the full set of parameters for the spatial profile is $\bm{\theta}^T = \{M_{500}, R_{500}, z, P_0, c_{500}, \gamma, \alpha, \beta\}$. {\corr We further divide these into two sets: $\bm{\theta}^S = \{P_0, c_{500}, \gamma, \alpha, \beta\}$ are the profile `shape' parameters, which may be universal, and $\bm{\theta}^P = \{M_{500}, R_{500}, z\}$ are the `physical' parameters, which are different for each cluster.}\footnote{{\corr The profile depends only on the combination $R_S = R_{500} / c_{500}$ (apart from in the mass scaling, $\alpha_p$), so these two parameters are degenerate for each cluster unless $c_{500}$ is assumed to be universal.}} The TSZ profile for a typical cluster is shown in Fig. \ref{fig:tsz-ksz-profile}, as a function of frequency and effective beam size.

For a sufficiently realistic model of the cluster spatial profile, there should be no need for a separate amplitude degree of freedom, $a_j$, because Eqs. (\ref{eqn-tsz-dt}) and (\ref{eqn-pe}) {\corr would} define a complete mapping between the magnitude of the TSZ signal, $y$, and physical cluster parameters such as $M_{500}$ and $R_{500}$. We introduce $a_j$ in our model for a few reasons, however. {\corr First of all, the mapping between the integrated SZ signal and parameters such as the mass is often defined using a scaling relation, which typically has an intrinsic scatter of $\log \sigma \sim 10\%$ \citep[e.g.][]{2011A&A...536A..11P}. The amplitude parameter can be used to model this scatter.

It is also} advantageous to be able to keep the cluster shapes fixed for some applications, since the spatial templates are time-consuming to compute. From Eq. (\ref{eqn-pe}), one can see that, {\corr assuming this profile is correct,} there is an almost one-to-one correspondence between $a_j$ and $M_{500}$, the parameter of most cosmological interest. The statistics of $a_j$ are also a good proxy for detection significance. By holding the shape parameters fixed, but allowing $a_j$ to vary, we can therefore get good estimates of these quantities with considerably reduced computational expense.

{\corr Finally, despite being one of the more accurate models available, the universal profile is still a simplification. \cite{2010A&A...517A..92A} find the pressure profiles of the REXCESS clusters to be scattered about the universal profile by up to a factor of four at low radii, depending on how morphologically disturbed the cluster is. The fit is claimed to be better than 25\% at radii greater than $0.2 \,R_{500}$ however, although recent SZ observations have now found a significantly flatter mean pressure profile at $r > R_{500}$ \citep{2013A&A...550A.131P, 2013ApJ...768..177S}.} Allowing $a_j$ to vary could at least help to reduce biases in other components of the data model due to this sort of modelling error (although one must be careful in choosing which profile parameters are also allowed to vary, as some are strongly degenerate with $a_j$).

\subsection{Amplitude sampling} \label{sec:tsz-amp-sampling}

Sampling of the $\{a_j\}$ parameters proceeds jointly with all other amplitude degrees of freedom, as described in Section \ref{sec:posterior-mapping}, but it is instructive to explicitly write out the linear system for just the CMB and a localised TSZ component, which we will now do.

From Section \ref{sec:posterior-mapping}, the data model for a single frequency may be written as $\d_\nu = \U_\nu \cdot \a + \n_\nu$, where
\bea
\a &=& (\a_\mathrm{CMB}, \; \a_\mathrm{TSZ}) \\
\U_\nu &=& \B_\nu (\bm{1}\cdot \bm{Y},\;  f(\nu) \T_\mathrm{TSZ})
\eea
are symbolic block vectors of the amplitudes and (beam convolved) mixing/projection operators for each component. The TSZ amplitudes are given by the block vector $\a_\mathrm{TSZ} = (a_1, a_2, \cdots, a_N)$, and the spatial templates by $\T_\mathrm{TSZ} = \T = (T_1, T_2, \cdots T_N)$, where $T_j = T(p, \hat{n}_j, \bm{\theta}_j)$.

In this notation, the linear operator (\ref{eqn-linsys-M}) can be written as $\bm{M} = \S^{-1} + \sum_\nu \tilde{\N}_\nu^{-1}$, with
\bea
\S^{-1} &=& \left( \begin{array}{ccc}
 \S^{-1}_\mathrm{CMB} \;  &     \\
   &  \; \S^{-1}_\mathrm{TSZ}  \end{array} \right) \nonumber \\
\tilde{\N}_\nu^{-1} &=& \left( \begin{array}{ccc}
B^T_\nu N^{-1}_\nu B_\nu   &   \;\;\; B^T_\nu N^{-1}_\nu B_\nu \T f(\nu)  \\
f(\nu) (B_\nu \T)^T N^{-1}_\nu B_\nu   &  \;\;\; f(\nu) (B_\nu \T)^T N^{-1}_\nu B_\nu \T f(\nu) \end{array} \right). \nonumber
\eea
We will set $\S^{-1}_\mathrm{TSZ} = 0$ in the rest of this paper, but include it here for the sake of generality. The bottom-right block of the inverse noise operator contains the TSZ-TSZ term
\be
(\tilde{\N}^{-1}_{\nu,22})_{jk} = f(\nu) (B_\nu T_j)^T N^{-1}_\nu B_\nu T_k f(\nu). \label{eqn-tszblock}
\ee
For $j \neq k$, this accounts for any overlap between clusters, ensuring that neighbouring clusters do not bias one another. Finally, the right-hand side of the linear system (\ref{eqn-linsys-b}) can be written as
\be
\bm{b} = \left( \begin{array}{ccc}
\S_\mathrm{CMB}^{-\frac{1}{2}} \omega_0 + \sum_\nu B_\nu^T \left [ N_\nu^{-1} d_\nu + N_\nu^{-\frac{1}{2}} \omega_\nu \right ] \\
\S_\mathrm{TSZ}^{-\frac{1}{2}} \omega_1 + \sum_\nu f(\nu) (B_\nu \T)^T \left [ N^{-1}_\nu d_\nu + N^{-\frac{1}{2}}_\nu \omega_\nu \right ]  \end{array} \right), \nonumber
\ee
where $\omega$ are randomly-drawn white noise maps.

In principle there is a physical prior on the TSZ amplitudes: $a_j \ge 0$. This violates our assumption that all amplitude parameters are Gaussian, without which we would be unable to simultaneously sample large numbers of amplitudes efficiently. To resolve this conflict, we choose a looser interpretation of $a_j$, treating it as a `diagnostic parameter' that quantifies detection significance, and is allowed to go negative.

%A significant detection of $a_j < 0$ for a cluster would signify a modelling error or similar; a negative value of $a_j$ that is consistent with zero would signify a non-detection (just as a positive value consistent with zero would). An alternative way around this problem would be to sample the cluster amplitudes in a separate Gibbs step, properly taking into account the prior; this would increase the correlation length of the Gibbs chain, however.

\subsection{{\corr Sampling the profile parameters}} \label{sec:tsz-shape-sampling}

The TSZ frequency spectrum is completely fixed, so there are no spectral parameters to sample. This leaves only the parameters of the {\corr spatial profile}, $\bm{\theta}^T$, defined above. {\corr The conditional distribution for $\bm{\theta}^T$ is difficult to sample from analytically due to the dependence of (\ref{eqn-tsz-template}) on a numerical line of sight integration, and the non-linear functional form of the pressure profile itself.} As such, we fall back on the Metropolis-Hastings algorithm to sample from Eq.~(\ref{eqn-shape-pdf}).\footnote{{\corr The shape parameter subspace, $\bm{\theta}^S$, could be sampled more directly by precomputing a set of profiles on a grid of the $\{\alpha, \beta, \gamma\}$ parameters and then rescaling the profiles with $R_S$, as necessary.}} This is tractable owing to the reasonably small number of {\corr profile} parameters for each cluster, although sampling is relatively slow because each proposal requires the cluster template to be recalculated for a new set of parameters. The parameters should strictly only be sampled one cluster at a time to preserve the Gibbs scheme, which disallows parallelisation of the sampling algorithm.

A couple of approximations can be made to speed up computations. The first involves assuming that all clusters share the same {\corr shape parameters, $\bm{\theta}^S$ (but not the same physical parameters, $\bm{\theta}^P$)}. In this case, the cluster profile recalculation can be parallelised effectively, and only one set of shape parameters need be sampled per Gibbs iteration. The shape parameters then represent some `average' profile for the ensemble of clusters.

The second involves approximating the likelihood for each cluster to be independent of the {\corr profile} parameters of any other clusters, in which case sampling for each cluster can happen in parallel. This is a good approximation unless clusters overlap. A possible refinement of this method would be to sample in parallel for all clusters except those that overlap by more than a pre-defined amount, for which sampling would instead happen sequentially.

\subsection{Catalogues and prior information} \label{sec:tsz-catalogue}

While recent high-resolution, high-sensitivity CMB experiments have greatly increased the number of clusters detected using the SZ effect, the majority are detected only with comparatively low SNR, or are barely resolved. The CMB data alone are therefore insufficient to strongly constrain the physical properties of most clusters, and we must look to other datasets to provide additional information. Fortunately, extensive cluster catalogues based on X-ray and galaxy redshift surveys are available (e.g. \cite{2007ApJ...660..239K, 2011A&A...534A.109P}), that can be used to put priors on {\corr some of} the cluster profile parameters, $\bm{\theta}^T$. Prior information is naturally incorporated into the Gibbs sampling procedure through the $P(\T)$ term in Eq.~(\ref{eqn-shape-pdf}).

Most important from our perspective are the redshift, characteristic mass ($M_{500}$), {\corr and scale radius ($R_S = R_{500} / c_{500}$) of the clusters}. Without some prior information on these parameters, the {\corr profile parameter} sampling method of Section \ref{sec:tsz-shape-sampling} {\corr can be affected by strong degeneracies, depending on exactly which set of parameters are being sampled}. Composite X-ray catalogues such as MCXC \citep{2011A&A...534A.109P} provide good estimates of these parameters for $\sim1800$ clusters, although it should be noted that the different selection functions for SZ and X-ray surveys mean that not all SZ clusters are present in the catalogue. Also, the parameters are typically estimated using scaling relations that are subject to systematic uncertainties in calibration, and which may disagree between SZ and X-ray observations \citep{2011A&A...536A..11P}.

A further requirement for the catalogues used by our Gibbs scheme is that they are free from duplicate entries. Eq. (\ref{eqn-tszblock}) shows why this is the case -- if cluster $k$ is actually a duplicate of $j$, there will be a 100\% overlap between them, leading to large off-diagonal entries in the TSZ-TSZ block of the linear operator. This can cause the system to become degenerate, leading to ill-defined solutions.

{\corr
\subsection{Contamination from compact sources} \label{sec:compact}

Several other source populations are known to contaminate the cluster TSZ signal. Infrared-emitting galaxies are a particularly nefarious contaminant, as they are often embedded in clusters but are typically not resolved by CMB experiments \citep{2012MNRAS.427.1741A, 2013JCAP...05..004H}, making it difficult to identify them and cleanly subtract their contribution to the signal. Cold galactic sources (CGS) in the Milky Way are also problematic, as they are sometimes found far from the galactic plane and have similar angular sizes to clusters \citep{2011A&A...536A..23P}, making them susceptible to erroneous identification as SZ sources at high frequencies, where the TSZ signal is a temperature increment.

While these objects are in principle distinguishable from the TSZ effect by their different spectra \citep[e.g.][]{2010A&A...522A..83M, 2014arXiv1409.6543A}, failing to account for them in the data model will result in some of their emission leaking into other components. This happens because the sampling algorithm has no way of identifying unmodelled components, and so simply tries to find the best fitting parameters of the incomplete model to the more complicated data, biasing the recovered SZ amplitudes and leaving residuals in other components such as the CMB map. Unless the contaminants can somehow be removed or masked, it is therefore necessary to have a sufficiently flexible component specification for contaminants of a signal, as well as the signal itself.

One possibility for dealing with this within our framework is to add a second localised component with the same spatial distribution as the clusters from the input catalogue, but with a different profile shape (e.g. the point source response in the case of unresolved galaxies) and frequency spectrum. One can then jointly sample the SZ and IR emission from each source, which would robustly separate the two contributions assuming a sufficient number of bands are available (auxiliary data from IR surveys such as IRAS, SCUBA, Herschel, and the high-frequency Planck HFI channels can also be used). This would require a reasonably informed choice of the functional form of the IR spectrum to prevent degeneracies with the SZ signal, but would have the advantage of simultaneously characterising the IR sources. One could also use this information to reliably distinguish cold galactic cores from true SZ clusters by means of a Bayesian model selection analysis,\footnote{{\corr We acknowledge the anonymous referee for this suggestion. Note that similar ideas have also been used to probabilistically distinguish between different supernova populations in contaminated Type Ia samples \citep{2010ApJ...723..398F}.}} although we do not consider this possibility further here.

}

%===============================================================================
\section{Kinetic SZ and peculiar velocities} \label{sec:ksz}

The kinetic SZ effect is also caused by the Compton scattering of CMB photons, but this time it is the coherent (bulk) motions of the scattering electrons with respect to the CMB that imprint the signal, which is effectively just a Doppler shift. This has a flat spectrum and is therefore not so readily distinguished from the primary CMB as the thermal SZ effect.

One can use the KSZ effect to probe the cosmological peculiar velocity field on large scales. This encodes a great deal of information about the growth of structure in the Universe, and can be used to constrain dark energy and modifications to General Relativity out to high redshift \citep{Bhattacharya:2007sk, Kosowsky:2009nc, 2013ApJ...765L..32K}. The KSZ effect is also sensitive to other phenomena that would cause a CMB dipole to be seen in the cluster rest frame, for example in inhomogeneous cosmological models that violate the Copernican Principle \citep{Goodman:1995dt, GarciaBellido:2008gd, Bull:2011wi}.

Because of the relative weakness of the signal and its lack of a distinctive spectral signature, the KSZ effect is susceptible to various systematic errors that can severely bias peculiar velocity measurements \citep{2001A&A...374....1A, Bhattacharya:2008qc}. The measured velocity is also degenerate with the optical depth of the cluster \citep{2005ApJ...635...22S}, so some way of independently determining this must be found. Our proposed approach is well-suited to addressing these difficult problems: through careful modelling, judicious use of prior information, and rigorous propagation of errors, one can break degeneracies and mitigate biases, even for extremely weak signals.

\subsection{Model definition}

For a cluster with bulk peculiar velocity $\bm{v}$, the fractional temperature change due to the KSZ effect is \citep{1980ARA&A..18..537S}
\bea \label{eqn-ksz-dt}
\frac{\Delta T}{T} &=& - (\bm{v} \cdot \hat{n}/c) \; \tau(\hat{n}) \\
 \tau(\hat{n}) &=& \int \sigma_T n_e(\hat{n}, l) dl.
\eea
The shape of the cluster's KSZ emission is governed by the electron number density, $n_e$. We already have a well-motived parametric form for the electron pressure profile (Eq. \ref{eqn-pe}), so rather than choosing $n_e$ independently, we use the ideal gas law $n_e \approx P_e / k_B T_e$ and a `universal' temperature profile \citep{2002ApJ...579..571L},
\be
T(r) = 11.2 \left (\frac{R_{500}h}{\mathrm{Mpc}} \right )^2 \left ( 1 + 0.75\frac{r}{R_{500}}\right )^{-1.6} ~\mathrm{keV},
\ee
to define an $n_e$ that is also a function of the TSZ shape parameters defined in Section \ref{sec:tsz-model}. By taking the TSZ and KSZ profiles to be governed by the same set of parameters, we explicitly take into account their common dependence on the physical properties of the cluster; information gleaned from the stronger TSZ signal helps break the degeneracy between the peculiar velocity and optical depth. The projection operator for the KSZ component is then
\be
T^\mathrm{KSZ}_j(p) = -\tau(p, \hat{n}_j, \bm{\theta}^T_j) / c, \label{eqn-ksz-template}
\ee
and the amplitude parameter is the bulk velocity projected along the line of sight to the cluster, $a_j = \bm{v}_j \cdot \hat{n}_j$. Since the KSZ effect is just a temperature change along the line of sight, it has a flat spectral dependence, so $\G = \bm{1}$.

\subsection{Velocity correlations} \label{sec:vcov}

In contrast with the TSZ case, we will not neglect the signal covariance here. The peculiar velocities of clusters are correlated over large scales, with a covariance that can be calculated from linear cosmological perturbation theory. While individual cluster velocities are of little intrinsic interest, velocity correlations can provide information on the growth of structure, matter power spectrum, and other cosmological parameters \citep{Bhattacharya:2006ke, Macaulay:2011av}.

The KSZ velocity covariance matrix is given by \citep{1988ApJ...332L...7G, Macaulay:2011av}
\be \label{eqn-vpec-correlation}
(S_\mathrm{KSZ})_{j k} = \int \frac{k^2 dk}{2 \pi^2} P_{vv}(k) F_{j k}(k) + \sigma^2_* \delta_{j k},
\ee
where the first term on the right hand side is the line of sight velocity correlation from linear theory, $\langle (\bm{v}_j\cdot \hat{n}_j) (\bm{v}_k\cdot \hat{n}_k) \rangle$, and the second term is a nonlinear velocity dispersion modelled as an uncorrelated noise term with variance $\sigma_*^2$. The window function $F_{j k}$ depends on the orientation of the pair of clusters $(j, k)$ with respect to the observer. Different (but equivalent) expressions for $F_{j k}$ are given by \cite{Dodelson:2003ft} and \cite{Ma:2010ps}. The velocity power spectrum is related to the matter power spectrum by
\bea
P_{vv}(k) &=& \kappa(z_j) \kappa(z_k) P(k, z=0)/k^2 \label{eqn-Pvv} \\
\kappa(z) &=& H(z) f(z) / (1 + z) D(z), \label{eqn-ksz-growth}
\eea
where $D(z)$ is the linear growth factor normalized to unity today and $f(z) = d \log D/d\log a$ is the growth rate.

\subsection{Velocity covariance matrix amplitude} \label{sec:vcov-amp}

Given a suitable choice of parametrisation, the cosmological functions that enter (\ref{eqn-Pvv}) can be constrained directly through the Gibbs sampling procedure. As a simple illustration, consider a parametrisation that has only the overall amplitude of the matter power spectrum, $A_\mathrm{PS}$, as a free parameter. Replacing $P(k) \mapsto A_\mathrm{PS} P_\mathrm{fid}(k)$ in (\ref{eqn-Pvv}), where $P_\mathrm{fid}(k)$ is a fiducial power spectrum, we can rewrite (\ref{eq:S_samp}) as
\bea
A_{\mathrm{PS}} & \leftarrow & P(A_\mathrm{PS} | \G, \a, \T, \d) \propto P(A_\mathrm{PS} | \a_\mathrm{KSZ}) \label{eqn-gibbs-aps} \\
  & \propto & \exp \left ( -\frac{1}{2} \mathbf{a}^\mathrm{T}_\mathrm{KSZ} \S^{-1}_\mathrm{KSZ} \mathbf{a}_\mathrm{KSZ} \right ) \left / \sqrt{|\S_\mathrm{KSZ}|} \right . . \label{eqn-gibbs-aps2}
\eea
If we set $\sigma_* = 0$, the signal covariance matrix is simply proportional to the fiducial linear theory velocity covariance matrix, $\S_\mathrm{KSZ} = A_\mathrm{PS} \S_\mathrm{fid}$, and the pdf reduces to the inverse gamma distribution,
\bea
\Gamma^{-1}(A; \alpha, \beta) &\propto& \exp ( -\beta / A) \; / \; A^{\alpha + 1}\\
 &\propto& \exp \left ( -\frac{1}{2} \mathbf{a}^\mathrm{T} \S_\mathrm{fid}^{-1} \mathbf{a} \;/\; A_\mathrm{PS} \right ) \left / \left (A^{N/2}_\mathrm{PS} \sqrt{|\S_\mathrm{fid}|} \right )\right ., \nonumber
\eea
where $N$ is the number of clusters and $\alpha = N/2 - 1$. Efficient direct sampling algorithms exist for this distribution \citep[e.g.][and references therein]{2004ApJS..155..227E}. We do not know of a direct sampler for the general case, where $\sigma_* \neq 0$, but an alternative method is discussed in Section \ref{sec:vcovsampler}.

%===============================================================================
\section{Gibbs sampler implementation} \label{sec:sampler}

Implementing a numerical code to efficiently carry out the sampling procedure described in previous sections is tractable but challenging. In this Section we discuss the computational difficulties associated with the proposed Gibbs sampling scheme, and suggest solutions for each of them. Our discussion is partially based on a simple proof-of-concept implementation for the TSZ effect built on top of the Commander CMB component separation code \citep{2004ApJS..155..227E, 2008ApJ...676...10E}. Commander already implements a subset of the proposed Gibbs scheme, and can be extended in a relatively modular fashion to accommodate a localised TSZ component. It is optimised for lower-resolution full-sky analyses, however, and lacks a suitable solver for the high-resolution analysis needed for TSZ clusters. We therefore use this code as a simple testbed, and defer full implementation to a later work.

\subsection{Full Gibbs scheme for SZ} \label{sec:implementation}

Combining results from the previous sections, a suitable Gibbs scheme for the localised TSZ and KSZ signals from galaxy clusters is
\begin{align}
\label{eq:gibbs_a}
\a^{i+1} &\leftarrow P(\a|C^i_\ell, A^i_\mathrm{PS}, \bm{\theta}^i_\mathrm{FG}, \bm{\theta}^i_\mathrm{SZ}, \d) 
\\\label{eq:gibbs_cl}
C^{i+1}_\ell &\leftarrow P(C_\ell|\a^{i+1}_\mathrm{CMB})
\\\label{eq:gibbs_aps}
A^{i+1}_\mathrm{PS} &\leftarrow P(A_\mathrm{PS}|\a^{i+1}_\mathrm{KSZ})
\\\label{eq:gibbs_theta_fg}
\bm{\theta}^{i+1}_\mathrm{FG} &\leftarrow P(\bm{\theta}_\mathrm{FG}|\a^{i+1}, \bm{\theta}^i_\mathrm{FG}, \bm{\theta}^i_\mathrm{SZ}, \d)
\\\label{eq:gibbs_theta_sz}
\bm{\theta}^{i+1}_\mathrm{SZ} &\leftarrow P(\bm{\theta}_\mathrm{SZ}|\a^{i+1}, \bm{\theta}^{i+1}_\mathrm{FG}, \bm{\theta}^i_\mathrm{SZ}, \d),
\end{align}
where $\a = (\a_\mathrm{CMB}, \a_\mathrm{FG}, \a_\mathrm{TSZ}, \a_\mathrm{KSZ})$ are the amplitude parameters, $\bm{\theta}_\mathrm{FG}$ are the foreground spectral parameters, and the cluster shape parameters $\bm{\theta}_\mathrm{SZ}$ are those of the universal pressure profile defined in Section \ref{sec:tsz-model}.

The Commander code already includes (\ref{eq:gibbs_a}), (\ref{eq:gibbs_cl}) and (\ref{eq:gibbs_theta_fg}). The amplitude sampling step (\ref{eq:gibbs_a}) must be generalised to include the new SZ components, and steps (\ref{eq:gibbs_aps}) and (\ref{eq:gibbs_theta_sz}) need to be implemented from scratch. We focus only on (\ref{eq:gibbs_a}) here, deferring detailed implementation of the other steps for later work.

\subsection{Constrained realisation solver} \label{sec:crsolver}

The computational complexity of the Gibbs scheme is entirely dominated by the joint amplitude sampling step (\ref{eq:gibbs_a}), which involves solving a large linear system to draw a constrained realisation of all of the amplitude parameters -- potentially millions of them. For realistic CMB data, with millions of multi-frequency pixels, inhomogeneous noise, and masked regions, there is a wide spread in the signal-to-noise ratio per pixel. The eigenvalues of the linear operator (\ref{eqn-linsys-M}) therefore have a large dynamic range, making the system poorly-conditioned. This, combined with its high dimensionality, results in unacceptably slow convergence for most linear solvers. Without a computationally-efficient global amplitude sampling step, the Gibbs scheme is intractable, so this issue is of central importance.

There are a number of ways to speed-up the solution of the linear system. Commander uses a preconditioned conjugate gradient (PCG) solver, which works by multiplying both sides of the system by a preconditioning matrix, $\mathbf{Q}$, and then solving the resulting modified system, $\mathbf{Q}\mathbf{M}\mathbf{a} = \mathbf{Q}\mathbf{b}$. If one can design a preconditioner such that $\mathbf{Q} \!\approx\! \mathbf{M}^{-1}$, the resulting modified system will be well-conditioned, and if both $\mathbf{Q}$ and $\mathbf{Q}\mathbf{M}$ can be evaluated quickly, it can be solved much faster. An efficient preconditioner for the joint amplitude sampling problem was described in \cite{2004ApJS..155..227E, 2008ApJ...676...10E}, and has been shown to work well on masked, full-sky, low-noise, multi-frequency foreground-contaminated CMB data up to $\ell \simeq 200$ \citep{2013arXiv1303.5072P}, which is sufficient for foreground component separation.

Higher resolution methods are needed to sample SZ amplitudes, however, as a typical cluster at $z \gtrsim 0.1$ subtends only a few arcminutes. Increasing the angular resolution by even a factor of 2-4 results in a considerable hardening of the problem, as the number of pixels required increases as the square. The noise is also higher at small scales, further contributing to the poor conditioning of the system. The result is that substantially more sophisticated solvers are required to make the problem tractable. One such method is the multi-level algorithm described by \cite{2014ApJS..210...24S}. This is capable of rapidly solving the amplitude sampling system up to $\ell \approx 2000$, but has yet to be extended to multi-frequency data with more than just the CMB plus noise.

An alternative is to reduce the complexity of the problem by working in the flat-sky limit. This is suitable for experiments such as ACT and SPT, which cover only a few thousand square degrees; so while their angular resolution is higher, the total number of pixels is typically smaller. Importantly, in the flat-sky limit one also benefits from being able to use Fast Fourier Transforms (FFTs) instead of the slower and more cumbersome spherical harmonic transforms.

\begin{figure}[t]
  \includegraphics[width=\columnwidth]{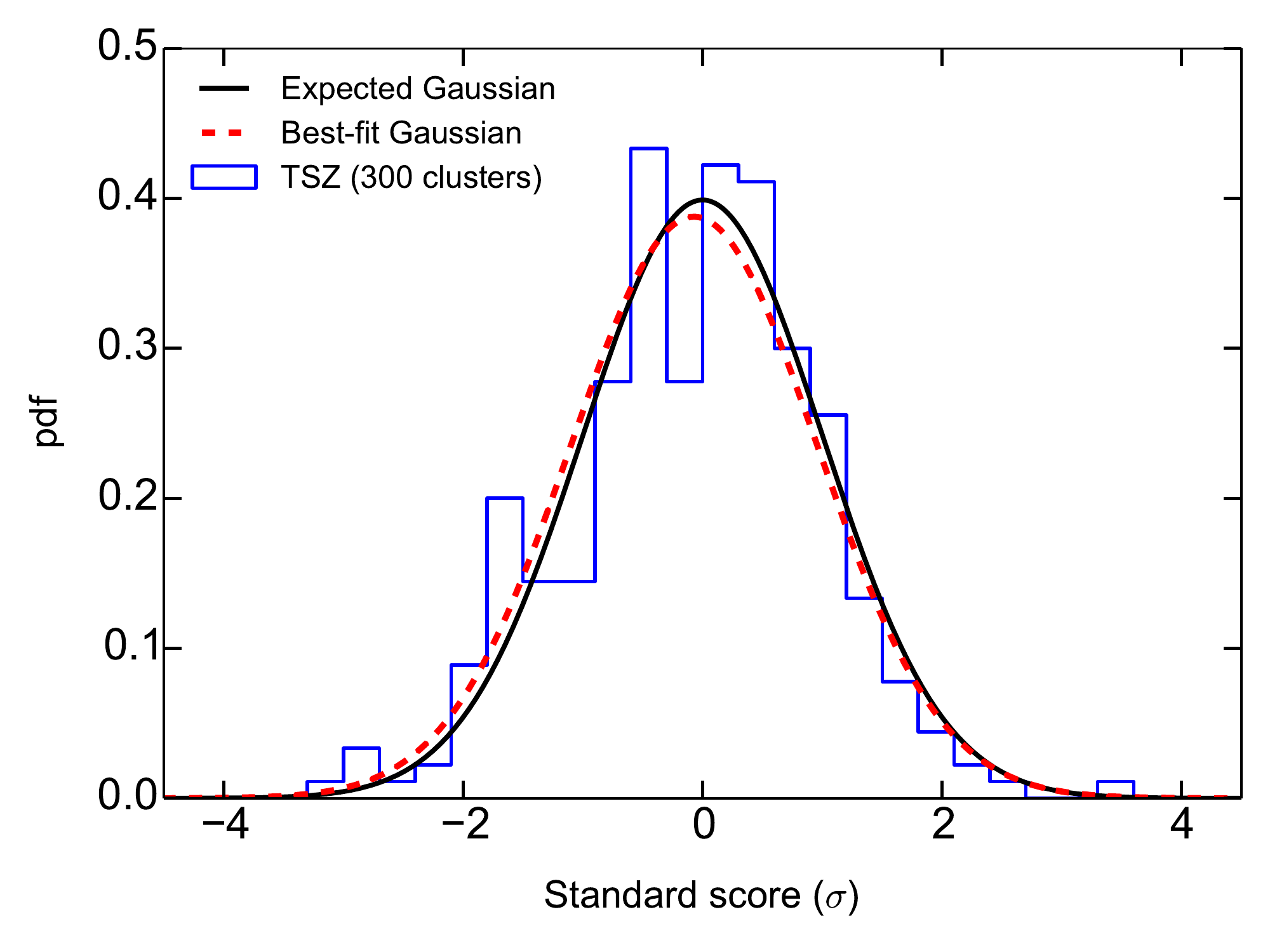}
  \caption{Distribution of standard scores, $(\langle a_i \rangle - a_{i,\mathrm{in}}) / \sigma_i$, for the TSZ amplitudes of 300 simulated clusters. The mean and standard deviation for each cluster are calculated from a Gibbs chain with 575 samples, and the input amplitudes are those used in the simulation.}
  \label{fig:zscores}
\end{figure}

To demonstrate the amplitude sampler, we apply the existing Commander PCG solver to a basic full-sky simulation for three frequency channels (143, 217, and 353 GHz), each of which has a primary CMB component, TSZ signals for 300 clusters, uncorrelated white noise, and instrumental beam effects. No foreground contamination or masks are included. A Gaussian CMB realisation is drawn using the Planck best-fit angular power spectrum \citep{2013arXiv1303.5075P}. The beam is chosen to be a uniform 40 arcmin. across all bands, and the noise covariance for each channel is obtained by smoothing the noise maps for the corresponding Planck HFI channels to the same resolution. To compensate for the comparatively low resolution of the simulation, we rescale the angular sizes of all clusters by a factor of 7, roughly corresponding to the ratio between the width of our chosen beam ($40'$) and the Planck HFI beams ($\sim 5'-7'$). Correspondingly, we scale the amplitudes of the clusters by $1/7^2$ in order to preserve their integrated flux, and thus the signal-to-noise ratio per cluster. The clusters are chosen to have the angular distribution and physical properties of the entries with the largest $\theta_{500}$ from the Planck SZ catalogue \citep{2013arXiv1303.5089P}, except Virgo and Coma which are too large after rescaling.

\begin{figure}[t]
  \hspace{-1em}\includegraphics[width=\columnwidth]{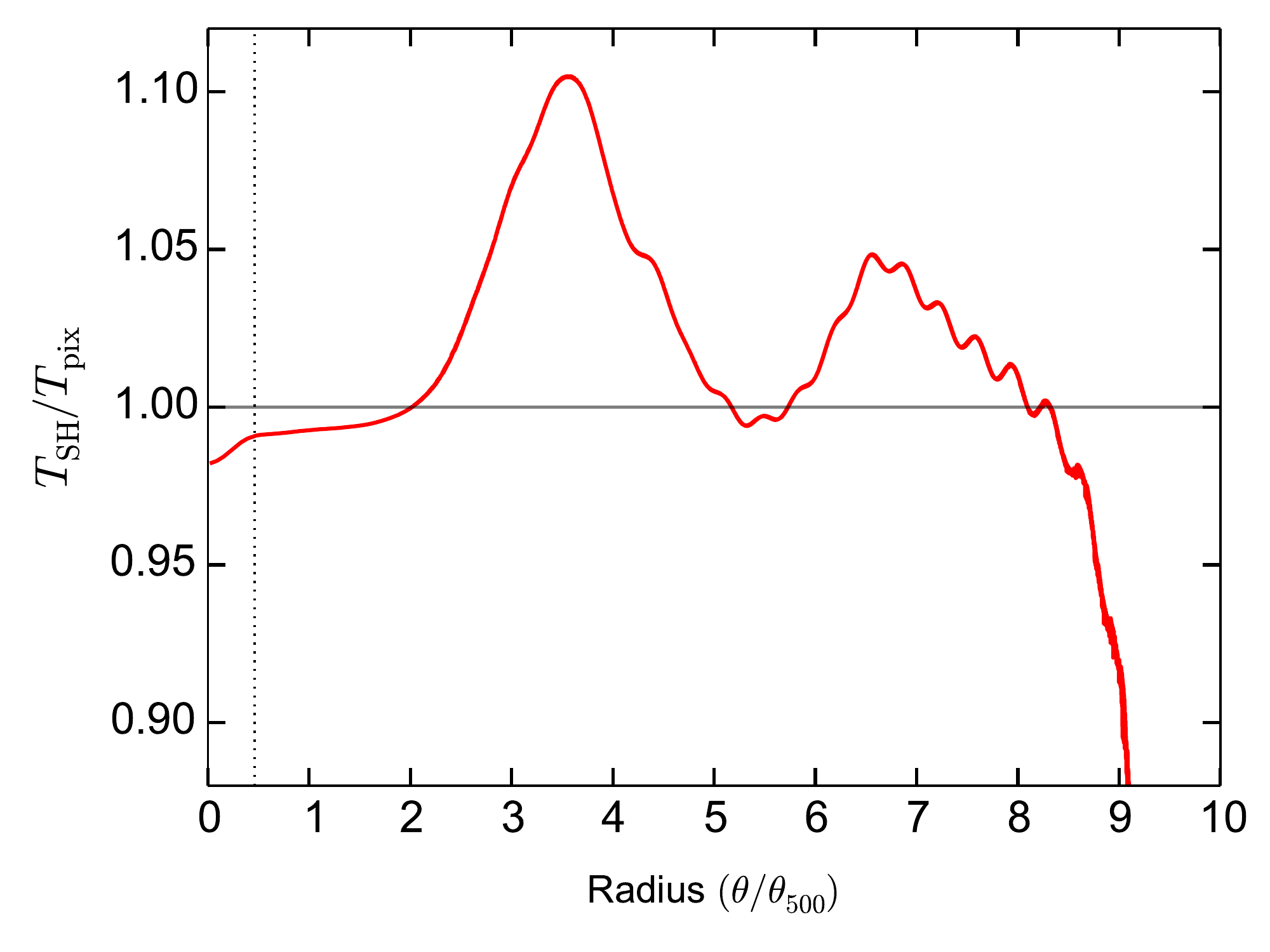}
  \caption{Ratio of TSZ templates for an example cluster ($\theta_{500} \simeq 30^\prime$) after beam convolution in spherical harmonic space and pixel space ($N_\mathrm{side}=1024$). The ratio is shown as a function of angle from the center of the cluster, normalised to $\theta_{500}$, and the beam FWHM ($\theta_\mathrm{FWHM} \simeq 14^\prime$) is shown as a dotted vertical line. Ringing artefacts are clearly visible.}
  \label{fig:beam-convol}
\end{figure}

Fig. \ref{fig:zscores} shows the distribution of cluster amplitudes recovered by running the modified Commander Gibbs sampler over the simulations. The standard scores (i.e. the recovered TSZ amplitude minus the amplitude input into the simulation, weighted by the standard deviation estimated from the Gibbs chain) are consistent with the unit Gaussian distribution. This is what one would expect if the recovered amplitudes are Gaussian-distributed and unbiased, and the standard deviation has been estimated correctly (in other words, that the statistical uncertainty has been propagated correctly). {\corr Note that this test has not yet been performed in the presence of other non-Gaussian foregrounds however, which will be required to more stringently validate the algorithm before it is applied to real CMB data.}

\subsection{Spatial template calculation}

The SZ spatial templates for each cluster are calculated according to Eqs. (\ref{eqn-tsz-dt}) and (\ref{eqn-ksz-dt}), both of which require line of sight integrations. This can be computationally intensive for high-resolution data, especially as the templates must be recalculated several times during the shape-sampling Gibbs step (\ref{eq:gibbs_theta_sz}). Simple numerical techniques like spline interpolation of the radial cluster profile and integrating for many clusters in parallel can readily be used to speed-up the process, however.

Once calculated, the templates must be convolved with the instrumental beam in each band. While the convolution would be fastest in spherical harmonic domain, this tends to introduce ringing artefacts, {\corr and can bias the beam-convolved template by a couple of percent at small radii, where most of the integrated signal is from (Fig. \ref{fig:beam-convol}). In turn, this biases the SZ amplitudes.} A much more accurate method is to perform the convolution directly on a finer grid in pixel space (i.e. each pixel is subdivided into 4 or 16 sub-pixels, the pixel-space convolution is calculated, and then the result is averaged back onto the coarser grid). This is considerably more expensive than the spherical harmonic method, so the beam-convolved templates should be cached if possible. Fortunately, the convolution can be done independently per cluster, per band, making it easy to parallelise.

\begin{figure}[t]
  \includegraphics[width=\columnwidth]{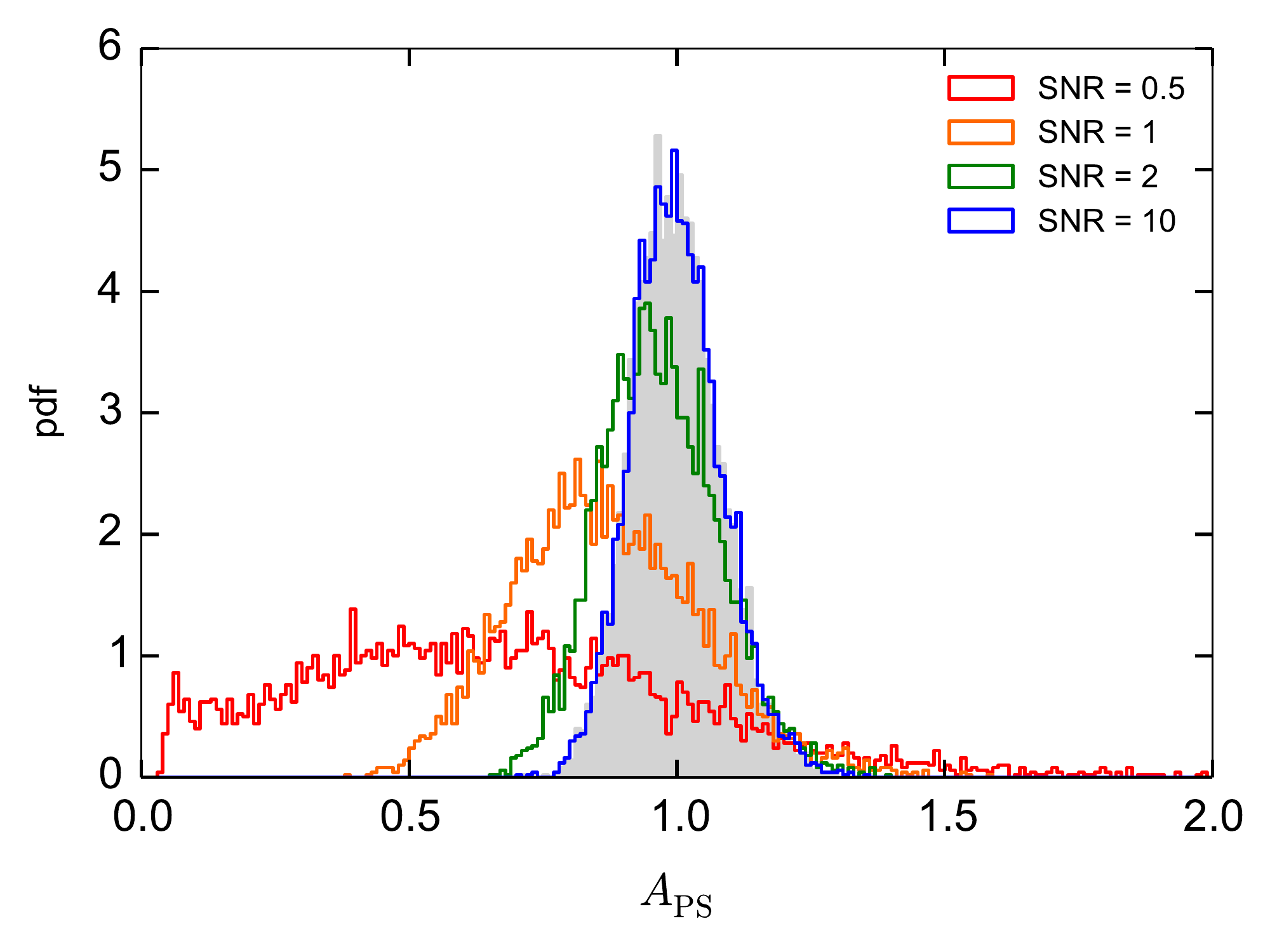}
  \caption{Marginal distributions of $A_\mathrm{PS}$ from simulations with different signal-to-noise ratios per cluster (for 300 clusters, 5000 samples). As the SNR increases, the marginal distribution converges towards the theoretical inverse gamma distribution for the given velocity covariance matrix (grey).}
  \label{fig:Aps}
\end{figure}

\subsection{Velocity covariance matrix amplitude} \label{sec:vcovsampler}

Current CMB experiments lack the sensitivity to detect the KSZ effect from significant numbers of individual galaxy clusters, so only statistical detections (e.g. \cite{2012PhRvL.109d1101H}) will be possible for the foreseeable future. Our formalism suggests a natural quantity to use as a statistic: the velocity covariance, defined in Section \ref{sec:vcov}.

To investigate the properties of this statistic, we define a simplified Gibbs scheme based on steps (\ref{eq:gibbs_a}) and (\ref{eq:gibbs_aps}), where only the cluster KSZ amplitudes and the velocity covariance matrix amplitude are free parameters,
\begin{align}
\a &\leftarrow P(\a|A_\mathrm{PS}, \d) \\
A_\mathrm{PS} &\leftarrow P(A_\mathrm{PS}|\a_\mathrm{KSZ}).
\end{align}
We simulate velocity data, $\mathbf{d}$, for a range of signal-to-noise ratios by drawing Gaussian realisations of velocities with covariance $\mathbf{S}_\mathrm{KSZ}$, and adding white noise with covariance $\mathbf{N} = (\mathrm{SNR})^{-2} \times \mathrm{diag}(\mathbf{S}_\mathrm{KSZ})$, where SNR is an assumed signal-to-noise ratio per cluster, equal for all clusters. The velocity covariance matrix is calculated for the 813 clusters with confirmed redshifts in the Planck SZ catalogue \citep{2013arXiv1303.5089P}; a subset of these is taken when fewer clusters are needed. For the sake of simplicity, the non-linear velocity dispersion, $\sigma_*$, is set to zero in these simulations; the qualitative picture stays the same for non-zero $\sigma_*$, however.

\begin{figure}[t]
  \includegraphics[width=\columnwidth]{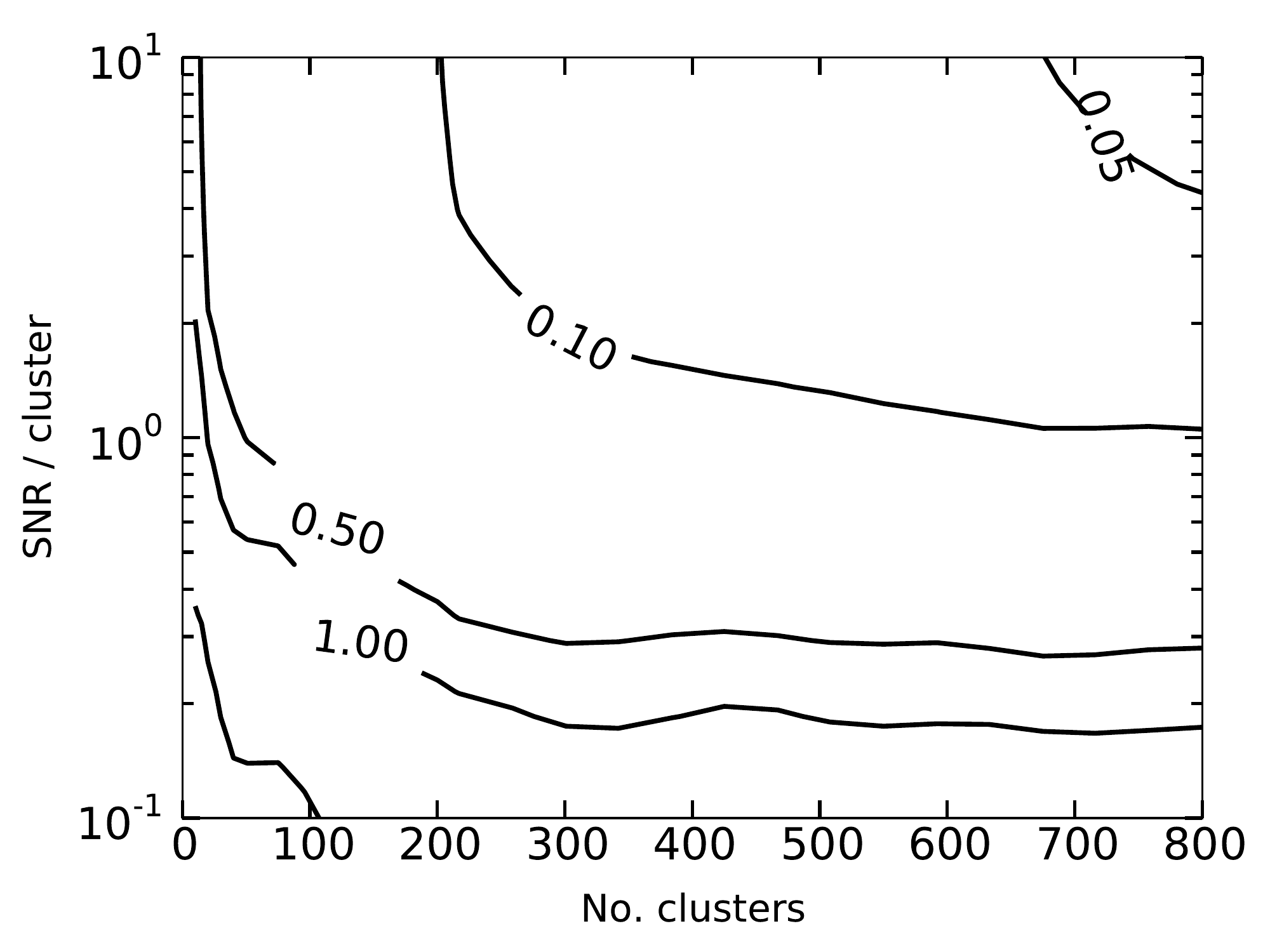}
  \caption{Standard deviation of the velocity covariance matrix amplitude, $\sigma(A_\mathrm{PS})$, estimated from simulations, as a function of the number of clusters in the catalogue and the detection SNR for each cluster.}
  \label{fig:vcoverr}
\end{figure}\vspace{1em}

One way of quantifying a detection of the KSZ effect is to use the amplitude parametrisation of the velocity covariance matrix that was discussed in Sect. \ref{sec:vcov-amp}. One might expect that a constraint on $A_\mathrm{PS}$ that is inconsistent with zero at some confidence level would count as a detection, but the velocity covariance matrix must remain positive definite, and so $A_\mathrm{PS}$ is always greater than zero with 100\% confidence. Furthermore, the distribution of $A_\mathrm{PS}$ will have finite variance even for perfect, noise-free observations -- a given (finite) set of correlated peculiar velocities is always consistent with having been drawn from a finite range of distributions with different $A_\mathrm{PS}$. Both of these effects can be seen in Fig. \ref{fig:Aps}, where the marginal distribution of $A_\mathrm{PS}$ is plotted from simulations for 300 clusters, with varying SNR. As the noise decreases, the marginal distribution rapidly approaches the ideal distribution for the given $\mathbf{S}_\mathrm{KSZ}$. The width of the estimated distribution, approximated by the standard deviation, is shown in Fig. \ref{fig:vcoverr} as a function of both SNR and the number of clusters.

A better way of quantifying detection significance is therefore to compare the estimated marginal distribution of $A_\mathrm{PS}$ with what would be expected in the ideal case. A useful measure of the closeness of two distributions is the Kullback-{\corr Leibler} divergence \citep{kullback1951}, also known as the relative entropy or information gain. For a pair of normalised discrete (binned) distributions, $p_i$ and $q_i$, this is
\be
\Delta S = \sum_i p_i \log(p_i / q_i),
\ee
where the sum is over bins. Since $\Delta S$ is not invariant under $q_i \leftrightarrow p_i$, we specify that $q_i$ is the ideal reference distribution here. As $p$ approaches $q$, $\Delta S \to 0$, so smaller $\Delta S$ denotes a stronger detection. Note that there is no definitive value of $\Delta S$ corresponding to a null detection, though; any constraint on the velocities, however weak, provides information on $A_\mathrm{PS}$, and thus reduces $\Delta S$.

In reality, $\sigma_* \neq 0$, and so Eq. (\ref{eqn-gibbs-aps2}) cannot be reduced to the inverse gamma distribution. One can sample from the more general distribution using a simple inversion sampling algorithm: (1) Evaluate Eq. (\ref{eqn-gibbs-aps2}) over a grid of $A_\mathrm{PS}$ values; (2) integrate the result to find the cumulative distribution function (cdf); (3) draw from the uniform distribution, $u \leftarrow \mathrm{U}[0, 1]$; and finally (4) draw $A_\mathrm{PS}$ by evaluating the (spline interpolated) inverse cdf at $u$, i.e., $A_\mathrm{PS} \leftarrow \mathrm{cdf}^{-1}(u)$. This method is sufficiently accurate as long as the $A_\mathrm{PS}$ grid is dense enough, and is also relatively efficient; while each evaluation of (\ref{eqn-gibbs-aps2}) requires an expensive matrix inversion and determinant evaluation, these are fast enough for matrices with a few thousand clusters or smaller, and the evaluation over the grid can be performed in parallel.

Another problem that arises is the long correlation length of Gibbs chains in the low SNR limit. In this case, the sampler spends most of its time exploring the joint prior, $P(\a_\mathrm{KSZ}, A_\mathrm{PS})$, and without good data to constrain the KSZ amplitudes, $A_\mathrm{PS}$ is degenerate with an overall scaling of all $\a_\mathrm{KSZ}$. Because the Gibbs scheme alternately samples from the conditional distributions, it is unable to move directly along the degeneracy direction, and so exploration of the joint prior is slow, hence the highly correlated samples. This issue can be overcome by adapting the low signal-to-noise CMB sampling algorithm of \cite{2009ApJ...697..258J}. This alternates between (1) sampling from the standard conditional distributions, and (2) an MCMC step based on a deterministic rescaling of the amplitudes. The MCMC step allows large jumps in $A_\mathrm{PS}$ in the noise-dominated regime, significantly speeding-up exploration of the joint prior space, but reduces to the standard sampling method in the signal-dominated case.

%===============================================================================
\section{Application to other localised signals} \label{sec:other}

The sampling scheme proposed in Section \ref{sec:gibbs} is suitable for a variety of other types of localised secondary anisotropy besides the TSZ and KSZ effects. The only significant modification that is needed is to choose a different parametric form for the spatial profile and frequency dependence, and to source a different catalogue of positions and (optionally) other properties of the target objects. In the remainder of this section, we briefly outline some other types of secondary anisotropy that may benefit from the careful statistical treatment provided by our method.

\subsection{Integrated Sachs-Wolfe effect} \label{sec:other-isw}

The integrated Sachs-Wolfe (ISW) effect is a temperature change in the CMB caused by the decay of gravitational potentials along the line of sight as dark energy begins to dominate. The expected amplitude of the ISW effect is predicted to be small compared with the primary anisotropies, and so it is generally necessary to cross-correlate CMB sky maps with tracers of large-scale structure in order to pick out the signal.

While interesting as an independent confirmation of the existence of dark energy, there have also been recent claims of a detection of the ISW effect in the direction of supervoids that is anomalously large compared to $\Lambda$CDM predictions \citep{2008ApJ...683L..99G, Papai:2010gd, Nadathur:2011iu, 2013JCAP...02..013F}. The ISW effect due to a large supervoid could potentially also explain the CMB cold spot \citep{2014arXiv1407.1470K}.

\vfill

\subsection{Topological defects} \label{sec:other-defects}

Phase transitions associated with spontaneous symmetry breaking in the early universe should give rise to topological defects -- the inhomogeneous boundaries between regions of different vacuum states \citep{Durrer:1999na}. Defects leave discontinuities and other characteristic non-Gaussian patterns in the CMB, with shapes that depend on the type of underlying symmetry that was broken. Most theories predict that only a few defects can be expected to be visible in the CMB, and so have been invoked as possible explanations of rare anomalies such as the cold spot \citep{Cruz:2004ce}.

There have been a number of searches for evidence of defects in the CMB, with varying degrees of success \citep{Jeong:2004ut, Cruz:2008sb, Feeney:2012jf}. In order to claim that a detected anomaly is in fact the result of a topological defect, a number of other potential explanations must first be ruled out; for example, a claimed detection of a cosmic texture may also be plausibly explained by the existence of an intervening cluster or void. The measured temperature profile of the anomaly can be used as part of a Bayesian model comparison to try and distinguish between the various options \citep{Cruz:2008sb}.

\subsection{Signatures of pre-inflationary physics} \label{sec:other-preinflation}

A number of proposed models of pre-inflationary physics suggest processes that can imprint patterns into the CMB that are not expected in the standard (Gaussian and isotropic) picture. These patterns typically take the form of circles, concentric rings, or other simple geometric shapes in CMB temperature or its variance, superimposed on anisotropies that are otherwise well described by a Gaussian random field. Examples of models which predict such effects include \citep{Wehus:2010pj}: bubble collisions in multiverse scenarios \citep{Feeney:2012hj}; cyclic cosmologies where primordial black holes collide \citep{Gurzadyan:2010da}; and models in which massive particles exist before inflation \citep{Fialkov:2009xm}. A closed spacetime topology would also give similar effects \citep{Cornish:1997ab}.

Such theories normally predict a characteristic shape for the patterns, and often impose other constraints such as pairing of the shapes, or fixed concentricities. This makes it possible to construct well-defined spatial profiles for the expected signal.

%===============================================================================
\section{Discussion} \label{sec:discussion}

Secondary anisotropies of the CMB are a rich source of cosmological information -- if they can be detected and characterised accurately. In this paper, we have described a Bayesian method to rigorously and reliably disentangle secondary signals in CMB temperature maps from other effects while simultaneously providing accurate estimates of statistical uncertainty. The basis of the method is a parametric physical model of the microwave sky that includes primary and secondary CMB anisotropies, foreground contamination, and noise. One can then use a tailor-made Gibbs sampling scheme to efficiently sample from the full joint posterior distribution for the model, which may include many thousands, or even millions, of parameters. After marginalising over everything else, one is left with a consistent statistical determination of the secondary signal. Though computationally intensive, this method has the key advantages of avoiding biases due to degeneracies with other signals (by modelling them), and correctly propagating uncertainties without relying on calibration against simulations or otherwise. The latter is particularly important for secondaries that are only marginally detected, where inaccuracies in error estimates could make the difference between claiming a detection or not.

While our proposed method guarantees statistical self-consistency, its ability to accurately describe the actual data depends on the suitability of the chosen model of the sky. As we discussed in Section \ref{sec:tsz}, the sampling framework is extremely flexible, supporting the sampling of cluster shape parameters and so on. It is therefore possible to arbitrarily extend the sky model to be more realistic, incorporating effects such as non-sphericity and non-thermal pressure support in galaxy clusters for example. Introducing additional Gibbs steps can lead to a significant increase in computational complexity, and generally increases uncertainties by requiring more parameters to be marginalised, so these downsides must be weighed against any expected improvement in accuracy.

In Section \ref{sec:ksz}, we gave an example of using our framework to directly estimate a cosmological statistic -- in this case, the velocity covariance. The advantage of this method is that the impact of foregrounds and the primary CMB can be folded directly into the estimated uncertainty on the statistic, based solely on the available data. The alternative is to calibrate the statistic off simulations, which can be computationally intensive and may lack some effects that are present in the real data. Mapping-out the full joint posterior distribution, as we do, also has other advantages; one can marginalise over the secondary signal to obtain rigorous estimates of some other signal component. This is of particular interest in cases where secondary anisotropies are both contaminants and interesting signals in their own right -- for example, the contamination of the cosmic infrared background by the thermal SZ effect. Though successful when applied to the CMB \citep{2013arXiv1303.5072P}, blind component separation methods are less likely to be useful for this sort of problem, as they tend to mix together physical foregrounds, leaving only the primary CMB behind after cleaning.

As discussed in Section \ref{sec:sampler}, our Gibbs sampling method is computationally intensive, and requires the use of some clever algorithms to speed it up. We demonstrated the tools necessary to make the method practical, but the next step is to construct a full implementation. For the full sky, the best option is most likely the multi-level solver of \cite{2014ApJS..210...24S}, but in the near term a flat sky version is a more straightforward prospect. Full details of a flat-sky implementation, including full numerical validation of the method, are deferred to a forthcoming paper \citep{LouisBull}.

The focus of this paper has been on secondary anisotropies of the CMB, such as the TSZ and KSZ effects, and the others listed in Section \ref{sec:other}, but one could also consider extending our Gibbs sampling framework to additional datasets, such as surveys of large scale structure (LSS). We have already considered a basic way of doing this in our discussion of the SZ effect, where external cluster catalogues were used to determine the positions of galaxy clusters on CMB maps. A more sophisticated `combined' Gibbs scheme could also sample parameters of the external cluster survey, such as detection thresholds or selection functions, or indeed any other quantity with statistical uncertainty attached. One could even define a scheme that sampled the density field reconstructed from galaxy redshift surveys and then cross-correlated it with the CMB. These possibilities are left for future work; we only wish to note here that properly accounting for correlated (physical/nuisance) parameters between disparate datasets is likely to become more crucial as cross-correlation analyses become more common, and that Gibbs sampling provides the flexibility to tackle this problem.

\acknowledgements {\it Acknowledgements ---} We are grateful to G. Addison, R. Battye, T. Louis, E. Macaulay, and M. Schammel for useful discussions, {\corr and to the anonymous referee for a number of suggestions that have substantially improved the paper}. This project was supported by ERC Starting Grant StG2010-257080. PB acknowledges additional support from STFC, and hospitality from Caltech/JPL. IKW acknowledges support from ERC grant 259505. PGF acknowledges support from Leverhulme, STFC, BIPAC and the Oxford Martin School. Part of the research was carried out at the Jet Propulsion Laboratory, California Institute of Technology, under a contract with NASA. Some of the results in this paper have been derived using the HEALPix software and analysis package \citep{Gorski:2004by}.

\appendix
\section{Blind detection of the TSZ effect} \label{app:blind}

The Gibbs sampling framework outlined in Section \ref{sec:gibbs} can also be used to perform blind detections of clusters, using only the frequency dependence of the thermal SZ effect. The most suitable model for blind TSZ detection within our framework is a pixel-based component (see Section \ref{sec:foregrounds}) with a free amplitude and fixed TSZ spectrum (Eq. \ref{eqn-tsz-spectrum}) in each pixel. One could also impose a signal covariance matrix based on the TSZ angular power spectrum predicted from linear theory and simulations, but for experiments with multiple frequency channels there is generally enough data to make this unnecessary.

While the spectrum of the TSZ effect is rather distinctive, other foreground emission must nevertheless be included in the data model as well. Otherwise, one runs the risk that a substantial fraction of the unmodelled components could be misidentified, causing significant contamination of the estimated TSZ signal. The precise definition of these other components will depend on the frequency coverage of the experiment in question, but for the 70 -- 350 GHz window, where the TSZ signal is largest, the most important contaminants are typically CO emission lines, free-free, and thermal dust \citep{2013arXiv1303.5072P}. Contamination by point sources is also an issue, and so either a point source mask or a reliable cross-matching procedure is required as well.

With a data model in hand, one can then use a variation on the Gibbs scheme of Section \ref{sec:posterior-mapping} to sample from the joint posterior. A map of TSZ amplitudes is produced with each Gibbs iteration by virtue of the first step in the scheme (Eq. \ref{eq:a_samp}). Once the Gibbs chain has converged, each of these maps will be a sample from the marginal distribution for the TSZ component -- that is, the CMB, galactic foregrounds, and other model parameters are automatically marginalised in these maps. This procedure directly propagates the uncertainty associated with the component separation procedure in full, so there is no need to perform simulations to estimate detection significance and the like.

The TSZ marginal maps can then be processed using an existing source finder or filtering technique to try and identify clusters \cite[e.g.][]{2005MNRAS.356..944H}. Estimates of the noise are given by the standard deviation from the chain for each pixel, and an effective signal map is given by the mean.

%===============================================================================
\bibliographystyle{hapj}
\bibliography{BibliographySZ}

\begin{thebibliography}{98}
\expandafter\ifx\csname natexlab\endcsname\relax\def\natexlab#1{#1}\fi

\bibitem[{{Addison} {et~al.}(2012){Addison}, {Dunkley}, \&
  {Spergel}}]{2012MNRAS.427.1741A}
{Addison}, G.~E., {Dunkley}, J., \& {Spergel}, D.~N. 2012, \mnras, 427, 1741,
  [arXiv:1204.5927]

\bibitem[{{Afshordi}(2004)}]{2004PhRvD..70h3536A}
{Afshordi}, N. 2004, \prd, 70, 083536, [arXiv:astro-ph/0401166]

\bibitem[{{Aghanim} {et~al.}(2001){Aghanim}, {G{\'o}rski}, \&
  {Puget}}]{2001A&A...374....1A}
{Aghanim}, N., {G{\'o}rski}, K.~M., \& {Puget}, J.-L. 2001, \aap, 374, 1,
  [arXiv:astro-ph/0105007]

\bibitem[{{Aghanim} {et~al.}(2014){Aghanim}, {Hurier}, {Diego}, {Douspis},
  {Macias-Perez}, {Pointecouteau}, {Comis}, {Arnaud}, \&
  {Montier}}]{2014arXiv1409.6543A}
{Aghanim}, N. {et~al.} 2014, ArXiv e-prints, [arXiv:1409.6543]

\bibitem[{{Aghanim} {et~al.}(2008){Aghanim}, {Majumdar}, \&
  {Silk}}]{2008RPPh...71f6902A}
{Aghanim}, N., {Majumdar}, S., \& {Silk}, J. 2008, Reports on Progress in
  Physics, 71, 066902, [arXiv:0711.0518]

\bibitem[{Allen {et~al.}(2011)Allen, Evrard, \& Mantz}]{Allen:2011zs}
Allen, S.~W., Evrard, A.~E., \& Mantz, A.~B. 2011, Ann.Rev.Astron.Astrophys.,
  49, 409, [arXiv:1103.4829]

\bibitem[{{Allison} \& {Dunkley}(2014)}]{2014MNRAS.437.3918A}
{Allison}, R., \& {Dunkley}, J. 2014, \mnras, 437, 3918, [arXiv:1308.2675]

\bibitem[{{Arnaud} {et~al.}(2010){Arnaud}, {Pratt}, {Piffaretti},
  {B{\"o}hringer}, {Croston}, \& {Pointecouteau}}]{2010A&A...517A..92A}
{Arnaud}, M., {Pratt}, G.~W., {Piffaretti}, R., {B{\"o}hringer}, H., {Croston},
  J.~H., \& {Pointecouteau}, E. 2010, \aap, 517, A92, [arXiv:0910.1234]

\bibitem[{{Atrio-Barandela} {et~al.}(2012){Atrio-Barandela}, {Kashlinsky},
  {Ebeling}, \& {Kocevski}}]{2012arXiv1211.4345A}
{Atrio-Barandela}, F., {Kashlinsky}, A., {Ebeling}, H., \& {Kocevski}, D. 2012,
  ArXiv e-prints, [arXiv:1211.4345]

\bibitem[{Battye \& Weller(2003)}]{Battye:2003bm}
Battye, R.~A., \& Weller, J. 2003, \prd, 68, 083506, [arXiv:astro-ph/0305568]

\bibitem[{Bhattacharya \& Kosowsky(2007)}]{Bhattacharya:2006ke}
Bhattacharya, S., \& Kosowsky, A. 2007, \apj, 659, L83,
  [arXiv:astro-ph/0612555]

\bibitem[{Bhattacharya \& Kosowsky(2008{\natexlab{a}})}]{Bhattacharya:2007sk}
------. 2008{\natexlab{a}}, \prd, 77, 083004, [arXiv:0712.0034]

\bibitem[{Bhattacharya \& Kosowsky(2008{\natexlab{b}})}]{Bhattacharya:2008qc}
------. 2008{\natexlab{b}}, JCAP, 0808, 030, [arXiv:0804.2494]

\bibitem[{{Birkinshaw} {et~al.}(1984){Birkinshaw}, {Gull}, \&
  {Hardebeck}}]{1984Natur.309...34B}
{Birkinshaw}, M., {Gull}, S.~F., \& {Hardebeck}, H. 1984, \nat, 309, 34

\bibitem[{{B{\"o}hringer} {et~al.}(2007){B{\"o}hringer}, {Schuecker}, {Pratt},
  {Arnaud}, {Ponman}, {Croston}, {Borgani}, {Bower}, {Briel}, {Collins},
  {Donahue}, {Forman}, {Finoguenov}, {Geller}, {Guzzo}, {Henry}, {Kneissl},
  {Mohr}, {Matsushita}, {Mullis}, {Ohashi}, {Pedersen}, {Pierini}, {Quintana},
  {Raychaudhury}, {Reiprich}, {Romer}, {Rosati}, {Sabirli}, {Temple}, {Viana},
  {Vikhlinin}, {Voit}, \& {Zhang}}]{2007A&A...469..363B}
{B{\"o}hringer}, H. {et~al.} 2007, \aap, 469, 363, [arXiv:astro-ph/0703553]

\bibitem[{Bull {et~al.}(2012)Bull, Clifton, \& Ferreira}]{Bull:2011wi}
Bull, P., Clifton, T., \& Ferreira, P.~G. 2012, \prd, 85, 024002,
  [arXiv:1108.2222]

\bibitem[{{Carvalho} {et~al.}(2012){Carvalho}, {Rocha}, {Hobson}, \&
  {Lasenby}}]{2012MNRAS.427.1384C}
{Carvalho}, P., {Rocha}, G., {Hobson}, M.~P., \& {Lasenby}, A. 2012, \mnras,
  427, 1384, [arXiv:1112.4886]

\bibitem[{Casella \& George(1992)}]{AmStat1992}
Casella, G., \& George, E.~I. 1992, The American Statistician, 46, pp. 167

\bibitem[{Cornish {et~al.}(1998)Cornish, Spergel, \& Starkman}]{Cornish:1997ab}
Cornish, N.~J., Spergel, D.~N., \& Starkman, G.~D. 1998, Class. Quant. Grav.,
  15, 2657, [arXiv:astro-ph/9801212]

\bibitem[{Cruz {et~al.}(2005)Cruz, Martinez-Gonzalez, Vielva, \&
  Cayon}]{Cruz:2004ce}
Cruz, M., Martinez-Gonzalez, E., Vielva, P., \& Cayon, L. 2005, \mnras, 356,
  29, [arXiv:astro-ph/0405341]

\bibitem[{{Cruz} {et~al.}(2008){Cruz}, {Mart{\'{\i}}nez-Gonz{\'a}lez},
  {Vielva}, {Diego}, {Hobson}, \& {Turok}}]{Cruz:2008sb}
{Cruz}, M., {Mart{\'{\i}}nez-Gonz{\'a}lez}, E., {Vielva}, P., {Diego}, J.~M.,
  {Hobson}, M., \& {Turok}, N. 2008, \mnras, 390, 913, [arXiv:0804.2904]

\bibitem[{{Das} {et~al.}(2011){Das}, {Sherwin}, {Aguirre}, {Appel}, {Bond},
  {Carvalho}, {Devlin}, {Dunkley}, {D{\"u}nner}, {Essinger-Hileman}, {Fowler},
  {Hajian}, {Halpern}, {Hasselfield}, {Hincks}, {Hlozek}, {Huffenberger},
  {Hughes}, {Irwin}, {Klein}, {Kosowsky}, {Lupton}, {Marriage}, {Marsden},
  {Menanteau}, {Moodley}, {Niemack}, {Nolta}, {Page}, {Parker}, {Reese},
  {Schmitt}, {Sehgal}, {Sievers}, {Spergel}, {Staggs}, {Swetz}, {Switzer},
  {Thornton}, {Visnjic}, \& {Wollack}}]{2011PhRvL.107b1301D}
{Das}, S. {et~al.} 2011, Phys. Rev. Lett., 107, 021301, [arXiv:1103.2124]

\bibitem[{{Diego} \& {Partridge}(2009)}]{2009arXiv0907.0233D}
{Diego}, J.~M., \& {Partridge}, B. 2009, arXiv e-prints, [arXiv:0907.0233]

\bibitem[{Dodelson(2003)}]{Dodelson:2003ft}
Dodelson, S. 2003, {Modern Cosmology} (Academic Press)

\bibitem[{Durrer(1999)}]{Durrer:1999na}
Durrer, R. 1999, New Astron.Rev., 43, 111

\bibitem[{{Eriksen} {et~al.}(2008){Eriksen}, {Jewell}, {Dickinson}, {Banday},
  {G{\'o}rski}, \& {Lawrence}}]{2008ApJ...676...10E}
{Eriksen}, H.~K., {Jewell}, J.~B., {Dickinson}, C., {Banday}, A.~J.,
  {G{\'o}rski}, K.~M., \& {Lawrence}, C.~R. 2008, \apj, 676, 10,
  [arXiv:0709.1058]

\bibitem[{{Eriksen} {et~al.}(2004){Eriksen}, {O'Dwyer}, {Jewell}, {Wandelt},
  {Larson}, {G{\'o}rski}, {Levin}, {Banday}, \& {Lilje}}]{2004ApJS..155..227E}
{Eriksen}, H.~K. {et~al.} 2004, \apjs, 155, 227, [arXiv:astro-ph/0407028]

\bibitem[{{Falck} {et~al.}(2010){Falck}, {Riess}, \&
  {Hlozek}}]{2010ApJ...723..398F}
{Falck}, B.~L., {Riess}, A.~G., \& {Hlozek}, R. 2010, \apj, 723, 398,
  [arXiv:1009.1903]

\bibitem[{Feeney {et~al.}(2013)Feeney, Johnson, McEwen, Mortlock, \&
  Peiris}]{Feeney:2012hj}
Feeney, S.~M., Johnson, M.~C., McEwen, J.~D., Mortlock, D.~J., \& Peiris, H.~V.
  2013, \prd, 88, 043012, [arXiv:1210.2725]

\bibitem[{Feeney {et~al.}(2012)Feeney, Johnson, Mortlock, \&
  Peiris}]{Feeney:2012jf}
Feeney, S.~M., Johnson, M.~C., Mortlock, D.~J., \& Peiris, H.~V. 2012, \prl,
  108, 241301, [arXiv:1203.1928]

\bibitem[{{Feroz} {et~al.}(2009){Feroz}, {Hobson}, {Zwart}, {Saunders}, \&
  {Grainge}}]{2009MNRAS.398.2049F}
{Feroz}, F., {Hobson}, M.~P., {Zwart}, J.~T.~L., {Saunders}, R.~D.~E., \&
  {Grainge}, K.~J.~B. 2009, \mnras, 398, 2049, [arXiv:0811.1199]

\bibitem[{Fialkov {et~al.}(2010)Fialkov, Itzhaki, \& Kovetz}]{Fialkov:2009xm}
Fialkov, A., Itzhaki, N., \& Kovetz, E.~D. 2010, JCAP, 1002, 004,
  [arXiv:0911.2100]

\bibitem[{{Flender} {et~al.}(2013){Flender}, {Hotchkiss}, \&
  {Nadathur}}]{2013JCAP...02..013F}
{Flender}, S., {Hotchkiss}, S., \& {Nadathur}, S. 2013, JCAP, 2, 13,
  [arXiv:1212.0776]

\bibitem[{Forni \& Aghanim(2005)}]{forni2005adapted}
Forni, O., \& Aghanim, N. 2005, EURASIP Journal on Applied Signal Processing,
  2005, 2413

\bibitem[{{Fosalba} {et~al.}(2003){Fosalba}, {Gazta{\~n}aga}, \&
  {Castander}}]{2003ApJ...597L..89F}
{Fosalba}, P., {Gazta{\~n}aga}, E., \& {Castander}, F.~J. 2003, \apjl, 597,
  L89, [arXiv:astro-ph/0307249]

\bibitem[{{Fowler} {et~al.}(2010){Fowler}, {Acquaviva}, {Ade}, {Aguirre},
  {Amiri}, {Appel}, {Barrientos}, {Battistelli}, {Bond}, {Brown}, {Burger},
  {Chervenak}, {Das}, {Devlin}, {Dicker}, {Doriese}, {Dunkley}, {D{\"u}nner},
  {Essinger-Hileman}, {Fisher}, {Hajian}, {Halpern}, {Hasselfield},
  {Hern{\'a}ndez-Monteagudo}, {Hilton}, {Hilton}, {Hincks}, {Hlozek},
  {Huffenberger}, {Hughes}, {Hughes}, {Infante}, {Irwin}, {Jimenez}, {Juin},
  {Kaul}, {Klein}, {Kosowsky}, {Lau}, {Limon}, {Lin}, {Lupton}, {Marriage},
  {Marsden}, {Martocci}, {Mauskopf}, {Menanteau}, {Moodley}, {Moseley},
  {Netterfield}, {Niemack}, {Nolta}, {Page}, {Parker}, {Partridge}, {Quintana},
  {Reid}, {Sehgal}, {Sievers}, {Spergel}, {Staggs}, {Swetz}, {Switzer},
  {Thornton}, {Trac}, {Tucker}, {Verde}, {Warne}, {Wilson}, {Wollack}, \&
  {Zhao}}]{2010ApJ...722.1148F}
{Fowler}, J.~W. {et~al.} 2010, \apj, 722, 1148, [arXiv:1001.2934]

\bibitem[{Garcia-Bellido \& Haugb{\o}lle(2008)}]{GarciaBellido:2008gd}
Garcia-Bellido, J., \& Haugb{\o}lle, T. 2008, JCAP, 0809, 016,
  [arXiv:0807.1326]

\bibitem[{Gelfand \& Smith(1990)}]{doi:10.1080/01621459.1990.10476213}
Gelfand, A.~E., \& Smith, A. F.~M. 1990, Journal of the American Statistical
  Association, 85, 398

\bibitem[{{Giannantonio} {et~al.}(2006){Giannantonio}, {Crittenden}, {Nichol},
  {Scranton}, {Richards}, {Myers}, {Brunner}, {Gray}, {Connolly}, \&
  {Schneider}}]{2006PhRvD..74f3520G}
{Giannantonio}, T. {et~al.} 2006, \prd, 74, 063520, [arXiv:astro-ph/0607572]

\bibitem[{Goodman(1995)}]{Goodman:1995dt}
Goodman, J. 1995, \prd, 52, 1821, [arXiv:astro-ph/9506068]

\bibitem[{{G\'{o}rski}(1988)}]{1988ApJ...332L...7G}
{G\'{o}rski}, K. 1988, \apjl, 332, L7

\bibitem[{G\'{o}rski {et~al.}(2005)G\'{o}rski, Hivon, Banday, Wandelt, Hansen,
  {et~al.}}]{Gorski:2004by}
G\'{o}rski, K., Hivon, E., Banday, A., Wandelt, B., Hansen, F., {et~al.} 2005,
  \apj, 622, 759, [arXiv:astro-ph/0409513]

\bibitem[{{Granett} {et~al.}(2008){Granett}, {Neyrinck}, \&
  {Szapudi}}]{2008ApJ...683L..99G}
{Granett}, B.~R., {Neyrinck}, M.~C., \& {Szapudi}, I. 2008, \apjl, 683, L99,
  [arXiv:0805.3695]

\bibitem[{Gurzadyan \& Penrose(2010)}]{Gurzadyan:2010da}
Gurzadyan, V.~G., \& Penrose, R. 2010, arXiv e-prints, [arXiv:1011.3706]

\bibitem[{{Hand} {et~al.}(2012){Hand}, {Addison}, {Aubourg}, {Battaglia},
  {Battistelli}, {Bizyaev}, {Bond}, {Brewington}, {Brinkmann}, {Brown}, {Das},
  {Dawson}, {Devlin}, {Dunkley}, {Dunner}, {Eisenstein}, {Fowler}, {Gralla},
  {Hajian}, {Halpern}, {Hilton}, {Hincks}, {Hlozek}, {Hughes}, {Infante},
  {Irwin}, {Kosowsky}, {Lin}, {Malanushenko}, {Malanushenko}, {Marriage},
  {Marsden}, {Menanteau}, {Moodley}, {Niemack}, {Nolta}, {Oravetz}, {Page},
  {Palanque-Delabrouille}, {Pan}, {Reese}, {Schlegel}, {Schneider}, {Sehgal},
  {Shelden}, {Sievers}, {Sif{\'o}n}, {Simmons}, {Snedden}, {Spergel}, {Staggs},
  {Swetz}, {Switzer}, {Trac}, {Weaver}, {Wollack}, {Yeche}, \&
  {Zunckel}}]{2012PhRvL.109d1101H}
{Hand}, N. {et~al.} 2012, \prl, 109, 041101, [arXiv:1203.4219]

\bibitem[{{Hern{\'a}ndez-Monteagudo}
  {et~al.}(2014{\natexlab{a}}){Hern{\'a}ndez-Monteagudo}, {Ross}, {Cuesta},
  {G{\'e}nova-Santos}, {Xia}, {Prada}, {Rossi}, {Neyrinck}, {Viel},
  {Rubi{\~n}o-Martin}, {Sc{\'o}ccola}, {Zhao}, {Schneider}, {Brownstein},
  {Thomas}, \& {Brinkmann}}]{Hernandez-Monteagudo:2013vwa}
{Hern{\'a}ndez-Monteagudo}, C. {et~al.} 2014{\natexlab{a}}, \mnras, 438, 1724,
  [arXiv:1303.4302]

\bibitem[{{Hern{\'a}ndez-Monteagudo}
  {et~al.}(2014{\natexlab{b}}){Hern{\'a}ndez-Monteagudo}, {Ross}, {Cuesta},
  {G{\'e}nova-Santos}, {Xia}, {Prada}, {Rossi}, {Neyrinck}, {Viel},
  {Rubi{\~n}o-Martin}, {Sc{\'o}ccola}, {Zhao}, {Schneider}, {Brownstein},
  {Thomas}, \& {Brinkmann}}]{2014MNRAS.438.1724H}
------. 2014{\natexlab{b}}, \mnras, 438, 1724, [arXiv:1303.4302]

\bibitem[{{Herranz} {et~al.}(2005){Herranz}, {Sanz}, {Barreiro}, \&
  {L{\'o}pez-Caniego}}]{2005MNRAS.356..944H}
{Herranz}, D., {Sanz}, J.~L., {Barreiro}, R.~B., \& {L{\'o}pez-Caniego}, M.
  2005, \mnras, 356, 944, [arXiv:astro-ph/0406226]

\bibitem[{{Herranz} {et~al.}(2002){Herranz}, {Sanz}, {Hobson}, {Barreiro},
  {Diego}, {Mart{\'{\i}}nez-Gonz{\'a}lez}, \& {Lasenby}}]{2002MNRAS.336.1057H}
{Herranz}, D., {Sanz}, J.~L., {Hobson}, M.~P., {Barreiro}, R.~B., {Diego},
  J.~M., {Mart{\'{\i}}nez-Gonz{\'a}lez}, E., \& {Lasenby}, A.~N. 2002, \mnras,
  336, 1057, [arXiv:astro-ph/0203486]

\bibitem[{{Hincks} {et~al.}(2013){Hincks}, {Hajian}, \&
  {Addison}}]{2013JCAP...05..004H}
{Hincks}, A.~D., {Hajian}, A., \& {Addison}, G.~E. 2013, JCAP, 5, 4,
  [arXiv:1303.3272]

\bibitem[{{Ho} {et~al.}(2008){Ho}, {Hirata}, {Padmanabhan}, {Seljak}, \&
  {Bahcall}}]{2008PhRvD..78d3519H}
{Ho}, S., {Hirata}, C., {Padmanabhan}, N., {Seljak}, U., \& {Bahcall}, N. 2008,
  \prd, 78, 043519, [arXiv:0801.0642]

\bibitem[{Jeong \& Smoot(2005)}]{Jeong:2004ut}
Jeong, E., \& Smoot, G.~F. 2005, \apj, 624, 21, [arXiv:astro-ph/0406432]

\bibitem[{Jewell {et~al.}(2004)Jewell, Levin, \& Anderson}]{Jewell:2002dz}
Jewell, J., Levin, S., \& Anderson, C. 2004, \apj, 609, 1,
  [arXiv:astro-ph/0209560]

\bibitem[{{Jewell} {et~al.}(2009){Jewell}, {Eriksen}, {Wandelt}, {O'Dwyer},
  {Huey}, \& {G{\'o}rski}}]{2009ApJ...697..258J}
{Jewell}, J.~B., {Eriksen}, H.~K., {Wandelt}, B.~D., {O'Dwyer}, I.~J., {Huey},
  G., \& {G{\'o}rski}, K.~M. 2009, \apj, 697, 258, [arXiv:0807.0624]

\bibitem[{{Keisler} \& {Schmidt}(2013)}]{2013ApJ...765L..32K}
{Keisler}, R., \& {Schmidt}, F. 2013, \apjl, 765, L32, [arXiv:1211.0668]

\bibitem[{{Koester} {et~al.}(2007){Koester}, {McKay}, {Annis}, {Wechsler},
  {Evrard}, {Bleem}, {Becker}, {Johnston}, {Sheldon}, {Nichol}, {Miller},
  {Scranton}, {Bahcall}, {Barentine}, {Brewington}, {Brinkmann}, {Harvanek},
  {Kleinman}, {Krzesinski}, {Long}, {Nitta}, {Schneider}, {Sneddin}, {Voges},
  \& {York}}]{2007ApJ...660..239K}
{Koester}, B.~P. {et~al.} 2007, \apj, 660, 239, [arXiv:astro-ph/0701265]

\bibitem[{Komatsu {et~al.}(2011)}]{Komatsu:2010fb}
Komatsu, E., {et~al.} 2011, \apjs, 192, 18, [arXiv:1001.4538]

\bibitem[{Kosowsky \& Bhattacharya(2009)}]{Kosowsky:2009nc}
Kosowsky, A., \& Bhattacharya, S. 2009, \prd, 80, 062003, [arXiv:0907.4202]

\bibitem[{{Kov{\'a}cs} {et~al.}(2014){Kov{\'a}cs}, {Szapudi}, {Granett},
  {Frei}, {Silk}, {Burgett}, {Cole}, {Draper}, {Farrow}, {Kaiser}, {Magnier},
  {Metcalfe}, {Morgan}, {Price}, {Tonry}, \& {Wainscoat}}]{2014arXiv1407.1470K}
{Kov{\'a}cs}, A. {et~al.} 2014, arXiv e-prints, [arXiv:1407.1470]

\bibitem[{Kullback \& Leibler(1951)}]{kullback1951}
Kullback, S., \& Leibler, R.~A. 1951, Ann. Math. Statist., 22, 79

\bibitem[{{Loken} {et~al.}(2002){Loken}, {Norman}, {Nelson}, {Burns}, {Bryan},
  \& {Motl}}]{2002ApJ...579..571L}
{Loken}, C., {Norman}, M.~L., {Nelson}, E., {Burns}, J., {Bryan}, G.~L., \&
  {Motl}, P. 2002, \apj, 579, 571, [arXiv:astro-ph/0207095]

\bibitem[{Louis \& Bull(2015)}]{LouisBull}
Louis, T., \& Bull, P. 2015, in preparation.

\bibitem[{Ma {et~al.}(2011)Ma, Gordon, \& Feldman}]{Ma:2010ps}
Ma, Y.-Z., Gordon, C., \& Feldman, H.~A. 2011, \prd, 83, 103002,
  [arXiv:1010.4276]

\bibitem[{Macaulay {et~al.}(2012)Macaulay, Feldman, Ferreira, Jaffe, Agarwal,
  {et~al.}}]{Macaulay:2011av}
Macaulay, E., Feldman, H.~A., Ferreira, P.~G., Jaffe, A.~H., Agarwal, S.,
  {et~al.} 2012, \mnras, 425, 1709, [arXiv:1111.3338]

\bibitem[{{Mak} {et~al.}(2011){Mak}, {Pierpaoli}, \&
  {Osborne}}]{2011ApJ...736..116M}
{Mak}, D.~S.~Y., {Pierpaoli}, E., \& {Osborne}, S.~J. 2011, \apj, 736, 116,
  [arXiv:1101.1581]

\bibitem[{{Matarrese} {et~al.}(2000){Matarrese}, {Verde}, \&
  {Jimenez}}]{2000ApJ...541...10M}
{Matarrese}, S., {Verde}, L., \& {Jimenez}, R. 2000, \apj, 541, 10,
  [arXiv:astro-ph/0001366]

\bibitem[{Melin {et~al.}(2012)Melin, Aghanim, Bartelmann, Bartlett, Betoule,
  {et~al.}}]{Melin:2012iz}
Melin, J.-B., Aghanim, N., Bartelmann, M., Bartlett, J., Betoule, M., {et~al.}
  2012, \aap, 548, A51, [arXiv:1210.1416]

\bibitem[{{Melin} {et~al.}(2006){Melin}, {Bartlett}, \&
  {Delabrouille}}]{2006A&A...459..341M}
{Melin}, J.-B., {Bartlett}, J.~G., \& {Delabrouille}, J. 2006, \aap, 459, 341,
  [arXiv:astro-ph/0602424]

\bibitem[{{Montier} {et~al.}(2010){Montier}, {Pelkonen}, {Juvela},
  {Ristorcelli}, \& {Marshall}}]{2010A&A...522A..83M}
{Montier}, L.~A., {Pelkonen}, V.-M., {Juvela}, M., {Ristorcelli}, I., \&
  {Marshall}, D.~J. 2010, \aap, 522, A83

\bibitem[{{Munshi} {et~al.}(2013){Munshi}, {Joudaki}, {Smidt}, {Coles}, \&
  {Kay}}]{2013MNRAS.429.1564M}
{Munshi}, D., {Joudaki}, S., {Smidt}, J., {Coles}, P., \& {Kay}, S.~T. 2013,
  \mnras, 429, 1564, [arXiv:1106.0766]

\bibitem[{Nadathur {et~al.}(2012)Nadathur, Hotchkiss, \&
  Sarkar}]{Nadathur:2011iu}
Nadathur, S., Hotchkiss, S., \& Sarkar, S. 2012, JCAP, 1206, 042,
  [arXiv:1109.4126]

\bibitem[{{Nagai} {et~al.}(2007){Nagai}, {Kravtsov}, \&
  {Vikhlinin}}]{2007ApJ...668....1N}
{Nagai}, D., {Kravtsov}, A.~V., \& {Vikhlinin}, A. 2007, \apj, 668, 1,
  [arXiv:astro-ph/0703661]

\bibitem[{Papai {et~al.}(2011)Papai, Szapudi, \& Granett}]{Papai:2010gd}
Papai, P., Szapudi, I., \& Granett, B.~R. 2011, \apj, 732, 27,
  [arXiv:1012.3750]

\bibitem[{{Pierpaoli} {et~al.}(2005){Pierpaoli}, {Anthoine}, {Huffenberger}, \&
  {Daubechies}}]{2005MNRAS.359..261P}
{Pierpaoli}, E., {Anthoine}, S., {Huffenberger}, K., \& {Daubechies}, I. 2005,
  \mnras, 359, 261, [arXiv:astro-ph/0412197]

\bibitem[{{Piffaretti} {et~al.}(2011){Piffaretti}, {Arnaud}, {Pratt},
  {Pointecouteau}, \& {Melin}}]{2011A&A...534A.109P}
{Piffaretti}, R., {Arnaud}, M., {Pratt}, G.~W., {Pointecouteau}, E., \&
  {Melin}, J.-B. 2011, \aap, 534, A109, [arXiv:1007.1916]

\bibitem[{{Planck Collaboration}(2011{\natexlab{a}})}]{2011A&A...536A..11P}
{Planck Collaboration}. 2011{\natexlab{a}}, \aap, 536, A11, [arXiv:1101.2026]

\bibitem[{{Planck Collaboration}(2011{\natexlab{b}})}]{2011A&A...536A..23P}
------. 2011{\natexlab{b}}, \aap, 536, A23, [arXiv:1101.2035]

\bibitem[{{Planck Collaboration}(2013)}]{2013A&A...550A.131P}
------. 2013, \aap, 550, A131, [arXiv:1207.4061]

\bibitem[{{Planck Collaboration}(2014{\natexlab{a}})}]{2013arXiv1303.5072P}
------. 2014{\natexlab{a}}, \aap, 571, A12, [arXiv:1303.5072]

\bibitem[{{Planck Collaboration}(2014{\natexlab{b}})}]{2013arXiv1303.5079P}
------. 2014{\natexlab{b}}, \aap, 571, A19, [arXiv:1303.5079]

\bibitem[{{Planck Collaboration}(2014{\natexlab{c}})}]{2013arXiv1303.5075P}
------. 2014{\natexlab{c}}, \aap, 571, A15, [arXiv:1303.5075]

\bibitem[{{Planck Collaboration}(2014{\natexlab{d}})}]{2013arXiv1303.5077P}
------. 2014{\natexlab{d}}, \aap, 571, A17, [arXiv:1303.5077]

\bibitem[{{Planck Collaboration}(2014{\natexlab{e}})}]{2013arXiv1303.5081P}
------. 2014{\natexlab{e}}, \aap, 571, A21, [arXiv:1303.5081]

\bibitem[{{Planck Collaboration}(2014{\natexlab{f}})}]{2013arXiv1303.5089P}
------. 2014{\natexlab{f}}, \aap, 571, A29, [arXiv:1303.5089]

\bibitem[{{Planck Collaboration}(2014{\natexlab{g}})}]{2014A&A...565A.103P}
------. 2014{\natexlab{g}}, \aap, 565, A103, [arXiv:1309.1357]

\bibitem[{{Rephaeli}(1995)}]{1995ARA&A..33..541R}
{Rephaeli}, Y. 1995, \araa, 33, 541

\bibitem[{{Sayers} {et~al.}(2013{\natexlab{a}}){Sayers}, {Czakon}, {Mantz},
  {Golwala}, {Ameglio}, {Downes}, {Koch}, {Lin}, {Maughan}, {Molnar},
  {Moustakas}, {Mroczkowski}, {Pierpaoli}, {Shitanishi}, {Siegel}, {Umetsu}, \&
  {Van der Pyl}}]{2013ApJ...768..177S}
{Sayers}, J. {et~al.} 2013{\natexlab{a}}, \apj, 768, 177, [arXiv:1211.1632]

\bibitem[{{Sayers} {et~al.}(2013{\natexlab{b}}){Sayers}, {Mroczkowski},
  {Zemcov}, {Korngut}, {Bock}, {Bulbul}, {Czakon}, {Egami}, {Golwala}, {Koch},
  {Lin}, {Mantz}, {Molnar}, {Moustakas}, {Pierpaoli}, {Rawle}, {Reese}, {Rex},
  {Shitanishi}, {Siegel}, \& {Umetsu}}]{2013ApJ...778...52S}
------. 2013{\natexlab{b}}, \apj, 778, 52, [arXiv:1312.3680]

\bibitem[{{Sch{\"a}fer} \& {Bartelmann}(2007)}]{2007MNRAS.377..253M}
{Sch{\"a}fer}, B.~M., \& {Bartelmann}, M. 2007, \mnras, 377, 253,
  [arXiv:astro-ph/0602406]

\bibitem[{{Sehgal} {et~al.}(2005){Sehgal}, {Kosowsky}, \&
  {Holder}}]{2005ApJ...635...22S}
{Sehgal}, N., {Kosowsky}, A., \& {Holder}, G. 2005, \apj, 635, 22,
  [arXiv:astro-ph/0504274]

\bibitem[{{Seljebotn} {et~al.}(2014){Seljebotn}, {Mardal}, {Jewell}, {Eriksen},
  \& {Bull}}]{2014ApJS..210...24S}
{Seljebotn}, D.~S., {Mardal}, K.-A., {Jewell}, J.~B., {Eriksen}, H.~K., \&
  {Bull}, P. 2014, \apjs, 210, 24, [arXiv:1308.5299]

\bibitem[{{Sherwin} {et~al.}(2012){Sherwin}, {Das}, {Hajian}, {Addison},
  {Bond}, {Crichton}, {Devlin}, {Dunkley}, {Gralla}, {Halpern}, {Hill},
  {Hincks}, {Hughes}, {Huffenberger}, {Hlozek}, {Kosowsky}, {Louis},
  {Marriage}, {Marsden}, {Menanteau}, {Moodley}, {Niemack}, {Page}, {Reese},
  {Sehgal}, {Sievers}, {Sif{\'o}n}, {Spergel}, {Staggs}, {Switzer}, \&
  {Wollack}}]{2012PhRvD..86h3006S}
{Sherwin}, B.~D. {et~al.} 2012, \prd, 86, 083006, [arXiv:1207.4543]

\bibitem[{{Shirokoff} {et~al.}(2011){Shirokoff}, {Reichardt}, {Shaw}, {Millea},
  {Ade}, {Aird}, {Benson}, {Bleem}, {Carlstrom}, {Chang}, {Cho}, {Crawford},
  {Crites}, {de Haan}, {Dobbs}, {Dudley}, {George}, {Halverson}, {Holder},
  {Holzapfel}, {Hrubes}, {Joy}, {Keisler}, {Knox}, {Lee}, {Leitch}, {Lueker},
  {Luong-Van}, {McMahon}, {Mehl}, {Meyer}, {Mohr}, {Montroy}, {Padin},
  {Plagge}, {Pryke}, {Ruhl}, {Schaffer}, {Spieler}, {Staniszewski}, {Stark},
  {Story}, {Vanderlinde}, {Vieira}, {Williamson}, \&
  {Zahn}}]{2011ApJ...736...61S}
{Shirokoff}, E. {et~al.} 2011, \apj, 736, 61, [arXiv:1012.4788]

\bibitem[{{Sunyaev} \& {Zeldovich}(1972)}]{1972CoASP...4..173S}
{Sunyaev}, R.~A., \& {Zeldovich}, Y.~B. 1972, Comments on Astrophysics and
  Space Physics, 4, 173

\bibitem[{{Sunyaev} \& {Zeldovich}(1980)}]{1980ARA&A..18..537S}
------. 1980, \araa, 18, 537

\bibitem[{Wandelt {et~al.}(2004)Wandelt, Larson, \&
  Lakshminarayanan}]{Wandelt:2003uk}
Wandelt, B.~D., Larson, D.~L., \& Lakshminarayanan, A. 2004, \prd, 70, 083511,
  [arXiv:astro-ph/0310080]

\bibitem[{Wehus \& Eriksen(2011)}]{Wehus:2010pj}
Wehus, I., \& Eriksen, H. 2011, \apj, 733, L29, [arXiv:1012.1268]

\bibitem[{{Wilson} {et~al.}(2012){Wilson}, {Sherwin}, {Hill}, {Addison},
  {Battaglia}, {Bond}, {Das}, {Devlin}, {Dunkley}, {D{\"u}nner}, {Fowler},
  {Gralla}, {Hajian}, {Halpern}, {Hilton}, {Hincks}, {Hlozek}, {Huffenberger},
  {Hughes}, {Kosowsky}, {Louis}, {Marriage}, {Marsden}, {Menanteau}, {Moodley},
  {Niemack}, {Nolta}, {Page}, {Partridge}, {Reese}, {Sehgal}, {Sievers},
  {Spergel}, {Staggs}, {Swetz}, {Switzer}, {Trac}, \&
  {Wollack}}]{2012PhRvD..86l2005W}
{Wilson}, M.~J. {et~al.} 2012, \prd, 86, 122005, [arXiv:1203.6633]

\end{thebibliography}

\end{document}